%% file: JASA_paper/JASA_submission_main.tex
\PassOptionsToPackage{unicode}{hyperref}
\PassOptionsToPackage{hyphens}{url}
\PassOptionsToPackage{dvipsnames,svgnames,x11names}{xcolor}
\documentclass[
  12pt]{article}

\usepackage{comment} %added
\usepackage{tikz} %added
\usetikzlibrary{arrows.meta,positioning,calc} %added
\usepackage{amsmath}   % provides align, gather, etc.
\usepackage{amssymb}   % additional symbols
\usepackage{amsfonts}  % defines \mathbb
\usepackage{amsthm}
\usepackage{mathtools} % fixes and extends amsmath
\usepackage{bbm} %added fp
\usepackage{subcaption}
\usepackage{enumerate}
\usepackage{natbib}
\usepackage{url} % not crucial - just used below for the URL 
\usepackage{booktabs}%added by xm 
\newtheorem{proposition}{Proposition} % ADDED BY XM

\usepackage{iftex}
\ifPDFTeX
  \usepackage[T1]{fontenc}
  \usepackage[utf8]{inputenc}
  \usepackage{textcomp} % provide euro and other symbols
\else % if luatex or xetex
  \usepackage{unicode-math}
  \defaultfontfeatures{Scale=MatchLowercase}
  \defaultfontfeatures[\rmfamily]{Ligatures=TeX,Scale=1}
\fi
\usepackage{lmodern}
\ifPDFTeX\else  
\fi
\IfFileExists{upquote.sty}{\usepackage{upquote}}{}
\IfFileExists{microtype.sty}{% use microtype if available
  \usepackage[]{microtype}
  \UseMicrotypeSet[protrusion]{basicmath} % disable protrusion for tt fonts
}{}
\makeatletter
\@ifundefined{KOMAClassName}{% if non-KOMA class
  \IfFileExists{parskip.sty}{%
    \usepackage{parskip}
  }{% else
    \setlength{\parindent}{0pt}
    \setlength{\parskip}{6pt plus 2pt minus 1pt}}
}{% if KOMA class
  \KOMAoptions{parskip=half}}
\makeatother
\usepackage{xcolor}
\makeatletter
\ifx\paragraph\undefined\else
  \let\oldparagraph\paragraph
  \renewcommand{\paragraph}{
    \@ifstar
      \xxxParagraphStar
      \xxxParagraphNoStar
  }
  \newcommand{\xxxParagraphStar}[1]{\oldparagraph*{#1}\mbox{}}
  \newcommand{\xxxParagraphNoStar}[1]{\oldparagraph{#1}\mbox{}}
\fi
\ifx\subparagraph\undefined\else
  \let\oldsubparagraph\subparagraph
  \renewcommand{\subparagraph}{
    \@ifstar
      \xxxSubParagraphStar
      \xxxSubParagraphNoStar
  }
  \newcommand{\xxxSubParagraphStar}[1]{\oldsubparagraph*{#1}\mbox{}}
  \newcommand{\xxxSubParagraphNoStar}[1]{\oldsubparagraph{#1}\mbox{}}
\fi
\makeatother

\usepackage{longtable,booktabs,array}
\usepackage{calc} % for calculating minipage widths
\usepackage{etoolbox}
\makeatletter
\patchcmd\longtable{\par}{\if@noskipsec\mbox{}\fi\par}{}{}
\makeatother
\IfFileExists{footnotehyper.sty}{\usepackage{footnotehyper}}{\usepackage{footnote}}
\makesavenoteenv{longtable}
\usepackage{graphicx}
\makeatletter
\def\maxwidth{\ifdim\Gin@nat@width>\linewidth\linewidth\else\Gin@nat@width\fi}
\def\maxheight{\ifdim\Gin@nat@height>\textheight\textheight\else\Gin@nat@height\fi}
\makeatother
\setkeys{Gin}{width=\maxwidth,height=\maxheight,keepaspectratio}
\makeatletter
\def\fps@figure{htbp}
\makeatother

\makeatletter
\@ifpackageloaded{caption}{}{\usepackage{caption}}
\AtBeginDocument{%
\ifdefined\contentsname
  \renewcommand*\contentsname{Table of contents}
\else
  \newcommand\contentsname{Table of contents}
\fi
\ifdefined\listfigurename
  \renewcommand*\listfigurename{List of Figures}
\else
  \newcommand\listfigurename{List of Figures}
\fi
\ifdefined\listtablename
  \renewcommand*\listtablename{List of Tables}
\else
  \newcommand\listtablename{List of Tables}
\fi
\ifdefined\figurename
  \renewcommand*\figurename{Figure}
\else
  \newcommand\figurename{Figure}
\fi
\ifdefined\tablename
  \renewcommand*\tablename{Table}
\else
  \newcommand\tablename{Table}
\fi
}
\@ifpackageloaded{float}{}{\usepackage{float}}
\floatstyle{ruled}
\@ifundefined{c@chapter}{\newfloat{codelisting}{h}{lop}}{\newfloat{codelisting}{h}{lop}[chapter]}
\floatname{codelisting}{Listing}

\makeatother
\makeatletter
\@ifpackageloaded{caption}{}{\usepackage{caption}}
\@ifpackageloaded{subcaption}{}{\usepackage{subcaption}}
\makeatother

\ifLuaTeX
  \usepackage{selnolig}  % disable illegal ligatures
\fi
\usepackage[]{natbib}
\usepackage{bookmark}

\IfFileExists{xurl.sty}{\usepackage{xurl}}{} % add URL line breaks if available
\hypersetup{
  pdftitle={Title},
  pdfauthor={Author 1; Author 2; Author 3},
  pdfkeywords={3 to 6 keywords, that do not appear in the title},
  colorlinks=true,
  linkcolor={blue},
  filecolor={Maroon},
  citecolor={Blue},
  urlcolor={Blue},
  pdfcreator={LaTeX via pandoc}}

\newcommand{\anon}{1}

\begin{document}

\def\spacingset#1{\renewcommand{\baselinestretch}%
{#1}\small\normalsize} \spacingset{1}

%%%%%%%%%%%%%%%%%%%%%%%%%%%%%%%%%%%%%%%%%%%%%%%%%%%%%%%%%%%%%%%%%%%%%%%%%%%%%%

 % \title{\bf Dynamic sparse graphs with  overlapping communities}
 % \author{
 %     Xenia Miscouridou$^{*$} \\
%    Department of Mathematics and Statistics, University of Cyprus \\Department of Mathematics, Imperial College London\\
 %   and \\
  % Francesca Panero$^{*$} \hspace{.2cm}\\ 
 %  Department of Methods and Models for Economics, Territory and Finance,\\
%   Sapienza University of Rome\\Department of Statistics, London School of Economics and Political Science
 %  and\\ 
%    Antreas Laos\hspace{.2cm}\\
%    Department of Computer Science, University of Cyprus\\
 %  }
 % \maketitle
  
\if1\anon
{
\title{Dynamic sparse graphs with  overlapping communities}

\author{Xenia Miscouridou$^{*,1,2}$, Francesca Panero$^{*,3,4}$, and Antreas Laos$^5$}

\date{
    \small
    $^1$Department of Mathematics and Statistics, University of Cyprus \\
    $^2$Department of Mathematics, Imperial College London\\
    $^3$Department of Methods and Models for Economics, Territory and Finance,\\
   Sapienza University of Rome\\
   $^4$Department of Statistics, London School of Economics and Political Science \\
    $^5$Department of Computer Science, University of Cyprus \\[1ex]
    $^*$These authors contributed equally to this work. 
    %\\ Corresponding author: \texttt{your.email@university.edu}
}
\maketitle
} \fi%

\if0\anon
{
  \bigskip
  \bigskip
  \bigskip
  \begin{center}
    {\LARGE\bf Dynamic sparse graphs with  overlapping communities}
\end{center}
  \medskip
} \fi

%$^{*}$ These authors contributed equally to this work.

\bigskip
\begin{abstract}
%Dynamic community detection in networks addresses the challenge of tracking how groups of interconnected nodes evolve, merge, and dissolve within time-evolving networks. Here, we propose a novel statistical framework for sparse networks with power-law degree distribution and dynamic overlapping community structure. Using a Bayesian Nonparametric framework, we build on the idea to represent the graph as an exchangeable point process on the plane. 
Dynamic community detection concerns inferring how community memberships evolve over time, including the emergence, persistence, merging, and dissolution of groups in temporal networks. We propose a Bayesian nonparametric model for time-evolving sparse networks, which captures power-law degree distributions and dynamically overlapping communities.
The model is constructed from vectors of completely random measures coupled through a latent Markov process governing the evolution of node affiliations. This construction provides a flexible and interpretable approach to model dynamic communities, naturally generalizing existing overlapping block models to the sparse and scale-free regimes.
We establish asymptotic results characterizing sparsity and degree heterogeneity over time, and develop an approximate inference procedure for recovering time-varying community trajectories. Applications to synthetic and real-world dynamic networks show that the model accurately uncovers evolving community structure and yields interpretable temporal patterns.

\end{abstract}

\noindent%
\vfill

\newpage
\spacingset{1.8} % DON'T change the spacing!

\section{Introduction}
\label{sec:intro}
Networks or graphs are mathematical structures that allow us to represent the relationships between a set of entities. These entities can be people, organizations, and they have various applications across the social sciences, biology, finance, economy and technology. For an overview see for example \cite{Newman2009} or \cite{Kolaczyk2009}. A generic network comprises of a set of nodes with edges between them and the degree of a node is the number of edges connected to it. The rate of growth of the number of edges in relation to the number of nodes defines the density (or sparsity) level of a graph. Intuitively, a dense network appears when the number of edges scales quadratically with the number of nodes, while a sparse network arises when this relation is subquadratic.
% Given the nature of these edges, we end up with binary graphs, multigraphs or weighted graphs. 

One of the central tasks in statistical network modeling is to be able to identify latent communities, i.e. groups of nodes that exhibit some sort of similar behavior and usually have comparable connectivity patterns. This is useful as real-world networks (social, biological, technological or others) often have modular structure~\citep{Newman2006,Fortunato2010}.
Community detection helps us uncover groups of nodes that are more densely connected internally than externally, revealing the hidden organization of the system. In social networks, communities may be social groups or interest circles. %In biology, they may represent protein complexes or functional gene modules. 
In information networks, communities identify related topics or content clusters (e.g., news or articles, research topics). In economics, they can highlight groups of markets or firms that interact closely.
%Various graph models with community structure (overlapping or not) exist in literature, just to name a few~\cite{Airoldi2008MMSB}, ~\cite{todeschini2020exchangeable}, {Miller et al., 2009;}. What is of particular interest however, is be able to model the dynamic behavior of such communities, i.e. to understand how these communities evolve in time. 

%Several models for static graphs exist, each admitting different properties, with the first graph of Erdos and Renyi (1959), to more recent approaches. For an overview, see eg Newman (2009). 
While static graph models have been widely applied, many real-world networks are inherently dynamic. The links between the nodes change in time as well as the community structure. 
For example, in a social network new connections appear or existing ones disappear, and similarly the communities are not static, as friend groups form, evolve, and dissolve over time showing us how social ties strengthen, weaken, or reorganize.
In information networks such as citation or news networks, dynamic communities uncover how topics appear, merge, or fade with time, which can shed light on the evolution of new fields, trends in online media, or shifting patterns of information diffusion.
%In a protein network proteins activate in different complexes over time, and 
%in a corporate network firms exhibit changes in their collaborative connection patterns in response to dynamic external reasons.
Therefore, in such settings a static analysis that aggregates the graph across time will obscure important features, such as the rate of turnover in membership, or the emergence of new groups. Hence, we need a class of dynamic communities network models that can characterize not only the structure of a network at a given moment, but also its evolutionary mechanisms such as whether communities are stable or transient, whether entities tend to remain in the same group or switch groups, and how global structural changes propagate through the system. Incorporating temporal dependence and correlation allows for information to be shared across adjacent time points rather than treating each timestep in isolation of the others. %For an overview of discrete time evolving networks see for example \cite{holme2015modern}. %check

Stochastic block models (SBMs) and their extensions have proven useful in statistics for modeling networks. Following the well celebrated static SBM, dynamic variations were introduced by~\cite{matias2017statistical,xu2013dynamic,Xu2012, yang2011detecting}. Although they have some differences (e.g. in inference) all these perform hard dynamic clustering on a graph.
%In~\cite{matias2017statistical} they propose a dynamic SBM with a Markov chain on the node labels. 
%Similar in spirit approaches are 
%~\cite{xu2013dynamic},~\cite{Xu2012}, 
%a state space evolution of SBM parameters takes place and in
%~\cite{Yang2011}.
%they use sliding window dynamics in all three approaches 
%though there is no overlapping communities but hard clustering of the nodes.
%as well as Riverain et al where in the latter they take a change point detection approach. 
In the static case,~\cite{Airoldi2008MMSB} proposed the mixed membership SBM. 
In the dynamic case~\cite{xing2010state} proposed a mixed membership model with nodes having `roles' and Logistic-normal role vectors with linear Gaussian state-space evolution. Inference is carried out through a variational Bayes expectation–maximization algorithm.~\cite{fu2009dynamic} is a previous and more concise version of~\cite{xing2010state}, and~\cite{ho2011evolving} also propose a dynamic mixed membership SBM. %. Similarly to~\cite{xing2010state}, propose a dynamic . %expand~\cite{xing2010state} by keeping a mixed membership approach with dynamic processes on clusters, rather than on the nodes. 
In similar spirit, other authors proposed dynamic models such as~\cite{Ishiguro2010}, who give a dynamic infinite relational model, and~\cite{herlau13} who deal with temporal networks of multiscale structure (vertices belong to a hierarchy of groups and subgroups). This list of papers is non-exhaustive, especially if one considers specific applications, e.g. see~\cite{martinet2020robust} who extract dynamic communities based on the explicit notion of time-evolving aggregations of smaller motifs on brain networks.
%Other dynamic network models exist that do not consider community structure.
%\textcolor{red}{Do we put the no communities dynamic graphs? Or the community static graphs}
Other network approaches which work on the latent space are~\cite{Hoff2009AMEN,Hoff01122002, sarkar2005dynlatent}, and~\cite{xu_zeng_2009}.
Beyond the task of community detection there are more time-varying graph models, such as the model for change point detection by~\cite{wilson2019modeling}. 
Similarly, if one goes beyond the scope of statistical models, current literature of deep learning on graphs exists. This is not within the scope of this paper, but for an overview on Graph Neural Networks see e.g.~\cite{Lingfei2022}. 

Classically, in a large number of papers, graphs were represented by the adjacency matrix of connections. At the same time, for both theoretical and practical reasons it was convenient to make an assumption of exchangeability. Under the adjacency matrix representation, exchangeability means that the graph's distribution is invariant to permutations of the nodes. However, Bayesian graphs that are represented by an
exchangeable random array are necessarily either empty or dense~\citep{Lovasz2013,orbanz2015bayesian}. This is also a persistent limitation of most existing dynamic SBMs, which is undesirable as many observed networks are sparse~\citep{Newman2009, orbanz2015bayesian}. Additionally, many sparse graphs exhibit heavy-tailed and power-law degree distributions, meaning that they have a few highly connected nodes and a large number of nodes with few connections (in particular, a large share has only one connection).

Combining exchangeability and sparsity is non-trivial. A solution was given by~\cite{Caron2017graph} who proposed an alternative way to model graphs that are distributionally invariant but range from dense to sparse, relying on a point process construction and its corresponding definition of exchangeability~\citep{Kallenberg1990}. A series of papers followed this construction~\citep{Borgs2018, Veitch2015, todeschini2020exchangeable, naik2022bayesian, herlau2014, Ricci2022thinnedrm}, none of which allows for dynamic communities. Here,  we introduce temporal dependence at the level of the completely random measures themselves, yielding a genuinely dynamic generative process rather than a sequence of conditionally independent static graphs. We propose a novel Bayesian nonparametric model for dynamic sparse graphs with overlapping communities (dynSNetOC)
%taking into consideration the above limitations. %combines the interpretability of blockmodels with the flexibility of Bayesian nonparametric approaches for sparse graph generation. Our model
which admits:\\
\indent 1. dynamic connectivity (with nodes and edges allowed to appear or disappear);\\
\indent 2. dynamic mixed membership community affiliations for each node;\\
\indent 3. sparsity and power-law degree distribution;\\
\indent 4. interpretable parameters and uncertainty quantification.\\
%Nodes are allowed to appear or disappear in subsequent timesteps and the model allows for node–community affiliations to evolve dynamically while simultaneously accommodating
%
Sparsity and power-law are achieved using exchangeable random measures. The dynamic community behavior arises from a latent Markov process and, as the community affiliations of the nodes vary, the connectivity also changes. Note that with the first point we also allow for nodes to enter later or exit earlier as opposed to other approaches which require the number of nodes to remain constant (e.g.~\cite{kang2022dynamic}). 
%Overall our model has interpretable parameters and

In Section~\ref{sec:model} we present the model, the latent Markov process, explain the graph properties, establish asymptotic results on sparsity and power-law and present a simulation algorithm. Section~\ref{sec:inference} describes a tractable approximation for the inference procedure, and in Section~\ref{sec:Experiments} we validate its use with synthetic data and then apply our method to a real-world graph. Empirical results show that dynSNetOC is able to recover the time-evolving communities while faithfully reproducing sparse connectivity and heavy-tailed degree distributions, while competitor models cannot capture all desirable properties. Indeed, ours is the only proposal able to accommodate all four characteristics together.
\section{Our model}
\label{sec:model}
%\label{sec:model}
\subsection{Representation of a graph as a point process}
\label{sec:Model_construction}
\cite{Caron2017graph} proposed an alternative framework for statistical network modelling, based on representing the graph as a point process. %~\cite{Todeschini2016} extended this to the multivariate version allowing for overlapping community affiliations. 
%Here, we extend the construction of~\cite{Todeschini2016} to a dynamic setup and present the statistical model for sparse, power-law multigraphs (with/out self-loops?) with dynamic mixed memberships. 
%~\cite{Todeschini2016} builds on vectors of CRMs~\cite{kingman1967} which we now make time dependent in order to allow for community memberships to evolve. % We provide here the necessary material for the definition of the network model; refer to section A of the on-line supplementary material for additional background on vectors of CRMs. Additionally, the model that is described in this section can be extended to bipartite graphs; see section G of the supplementary material for more details.
%\begin{itemize}
    %\item
    %\subsubsection{Static simple graph}
    %The model is based on a point process construction. % (but multigraph, poisson counts) and given some parameters they have bernoulli links.  
Specifically, a graph is modeled as an exchangeable
random measure on the plane where each node is embedded at
some location $\theta_{i}\in\mathbb{R}_{+}$. %and, for simple graphs, a connection exists between
%two nodes $i$ and $j$ if there exists a point at locations $(\theta_{i},\theta_{j})$ and
%$(\theta_{j},\theta_{i})$. 
A directed multigraph is represented by an atomic measure on the plane
\begin{equation*}
N=\sum_{i,j}n_{ij}\delta_{(\theta_{i},\theta_{j})},\label{eq:pointprocessZ}%
\end{equation*}
where $n_{ij}$ is the number of directed edges from node $i$ to $j$. % if $i$ and $j$ are connected, 0 otherwise.%; see Figure~\ref{fig:pointprocess} for an illustration. $z_{ij}$ 
%They consider the following simple generative model, where two nodes $i\neq j$ connect with probability
The links are Poisson distributed
\begin{equation*}
n_{ij}|(w_{\ell })_{\ell=1,2,\ldots}\sim \text{Poisson}(w_{i}w_{j}),\qquad i\neq j \label{eq:link1}%
\end{equation*}
%where the $(w_i,\theta_i)_{i=1,2,\ldots}$ are the points of a Poisson
%point process on $\mathbb{R}_{+}^{2}$. %For (symmetric) multigraphs the point process cosntruction would be \begin{equation}
%N=\sum_{i,j}n_{ij}\delta_{(\theta_{i},\theta_{j})},\label{eq:pointprocessN}
%\end{equation} 
%where $n_ij$ is the number of links (interactions) between i and j.
where the weights $(w_i)$ represent node latent variables that make it more or less prone to form connections. The weights and locations
$(w_{i}, \theta_i)$ are draws from a Poisson process on $(0,\infty)\times \mathbb{R}_+$ defined by a mean measure
$\nu(dw, d\theta)=\rho(dw) \lambda(d\theta),$ where $\lambda$ is the Lebesgue measure and $\rho$ is a $\sigma$-finite measure 
on $\mathbb{R}_{+}$, concentrated on $\mathbb{R}_{+} \setminus \{0\}$, which satisfies
$
\int_{\mathbb{R}_{+}} \min \!\left( 1, w \right)
\rho(dw) < \infty.$
Under the prescribed conditions, the weights and locations ($w_{i},
\theta_i$) are described by a completely random measure (CRM)~\citep{Kingman1993Poisson} on $\mathbb{R}_+$:
\[
W=\sum_{i=1}^\infty w_{i}\delta_{\theta_i},\qquad W\sim \text{CRM}(\rho,\lambda).
 \label{eq:CRM}
\]
 More on the CRM construction and the properties of exchangeability are given in~\cite{Caron2017graph}. 
Then, a directed graph $N$ is generated from a Poisson process %with intensity given by the product measure $W\times W$ on $\mathbb{R}$,  
$N\mid W\sim\text{Poisson}\left(W \times W\right).$
%\textcolor{red}{F: I would not put the GGP here but only in our model. I would just write that the activity of the process implies the sparsity and density.} 
The properties of the L\'evy measure $\rho$ will define the sparsity/density of the graph and other asymptotic properties of the graph. Therefore, its choice will be pivotal to the graph construction, as we will prove in section \ref{subsec:asymptotics}. If $\int_0^\infty \rho(dw) =\infty$, there are an infinite number of jumps in any finite interval and the CRM is said to have infinite activity. %In this case, the number of edges scales subquadratically with the number of observed nodes and the resulting graph is sparse, i.e. its edges grow subquadratically with the number of nodes. 
Otherwise, the number of jumps is finite almost surely and the CRM has finite activity.% and resulting graph is dense (the number of edges scales quadratically with the number of observed nodes). Note that sparsity is an asymptotic property and it is reasonable to talk about it only with a large number of nodes.

%The L\'{e}vy measure is taken to be a Generalized Gamma Process, i.e.
%\begin{equation}
%\rho_0(dw_0)=\frac{1}{\Gamma(1-\sigma)}w_0^{-1-%\sigma} e^{-w_0 \tau}dw_0
%\label{eq:L\'evyGGP}
%\end{equation}
%for $\sigma \in (\infty, 1), \tau>0, \alpha>0$. We write GGP($\alpha, \sigma, \tau$). If $\sigma<0$ then the GPP is finite activite and the resulting graph is almost surely dense whereas if $\sigma \in (0,1)$ the GGP is of infinite activity and the resulting graph is almost surely sparse, i.e. its edges grow subquadratically with the number of nodes. For more on sparsity see ~\cite{Caron2014a, caron2017sparsity}.
%ADD A PICTURE OF THE POINT PROCESS??\textcolor{red}{I think one is sufficient, if we want we can do it for the dynamic version}\\
%
%
\subsection{Construction}
\label{sec:construction}
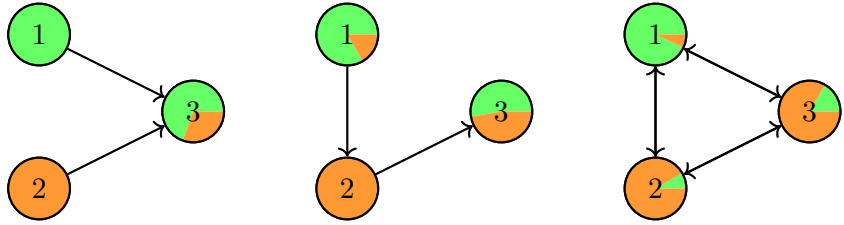
\begin{figure}[t!]
\begin{center}
\centering
{\hspace{-0.15in}\resizebox{0.7\textwidth}{!}{\input{JASA_paper/n_2.tex}}}
%{\includegraphics[width=.48\textwidth]{samplegraph3.png}}
\end{center}
\caption{Snapshots of a graph with $3$ nodes and 3 timesteps. There are $2$ latent communities (green and orange), and each node is colored according to its affiliations.} %Loosely speaking for $t=1,2$ nodes $1,2,3$ have more links between them whereas node $4$ is less connected with them. This explains why nodes $1,2,3$ belong to the green community whereas node $4$ belongs to the orange one for $t=1,2$. However at $t=3$ nodes $2$ and $3$ join the orange community. This aligns with the fact that at $t=3$ there are more links between $2,3,4$ whereas node $1$ is more `isolated'.}
%\caption{Illustration of the snapshots of a toy example for a graph with $4$ nodes for timesteps $t=1,2,3$. In the current time window we assume that there are $2$ latent communities (green and orange), and color each node according to its highest affiliation. At each timestep new edges appear or disappear between the nodes. Loosely speaking for $t=1,2$ nodes $1,2,3$ have more links between them whereas node $4$ is less connected with them. This explains why nodes $1,2,3$ belong to the green community whereas node $4$ belongs to the orange one for $t=1,2$. However at $t=3$ nodes $2$ and $3$ join the orange community. This aligns with the fact that at $t=3$ there are more links between $2,3,4$ whereas node $1$ is more `isolated'.}
\label{fig:toyexample}
\end{figure}
Our proposed model dynSNetOC extends the overlapping community model of~\cite{todeschini2020exchangeable} to the
case where the nodes' affiliation scores to the communities evolve over time following a latent Markov process. 
In this way, we generalize the dynamic mixed membership stochastic blockmodel model to the sparse and power-law regime. Before giving a mathematical formulation, we demonstrate graphically our model. Figure~\ref{fig:toyexample} illustrates a toy example of our model with $3$ nodes, $3$ timesteps and $2$ communities, shown in green and orange. The arrows between the nodes show the directed links from one node to another and, as shown, these links change in time. The affiliations are shown in pie charts in green and orange corresponding to the proportion of affiliation to the two communities. At each timestep new edges appear or disappear between the nodes and the affiliations change too. %Node $1$ is green at $t=1$ and although some orange proportion appears, the majority remains green for $t=2,3$. Node $2$ is clearly orange in all timesteps. Node $3$ is firstly mostly green, although with an orange proportion too. Then, this increases and by $t=3$ it is mostly orange and will be classified as orange. %%%%Note that the scores $\beta_{ti}$ for a node $i$ at time $t$ do not sum up to 1 but we can normalize them for illustration and interpretation purposes as done in Figure~\ref{fig:toyexample}.

We now formally introduce the model, while the interpretation of the parameters is postponed to Section~\ref{subsec:interpretation}.
We model graphs with $p$ overlapping communities using vectors of completely random measures which evolve in time. %These evolv over the timesteps $t=1,\dots,T$ we % initially proposed in~\cite{Griffin2014}
%(or equivalently multivariate subordinators) and extend it to vary over time. 
%At each timestep $t$, the process is marginally similar to the construction of~\cite{Caron2017graph}, but extended to consider $p$ communities through a vector of $p$ CRMs.
We define $(W^{(t)}_1,\ldots,W^{(t)}_p) \text{ on } \mathbb R_+^p$, where each component is a homogeneous CRM with independent increments
$W^{(t)}_k=\sum_{i=1}^\infty w^{(t)}_{ik}\delta_{\theta_i}.$
%and write
%\begin{equation}
%\left(W^{(t)}_1,\ldots,W^{(t)}_p\right)\sim \text{CRM}\left(\rho^{(t)},\lambda\right).\label{eq:vCRM}
%\end{equation}
The dynamic multigraph is thus described as follows:
\begin{align}
\begin{aligned} \begin{array}{ll} W^{(t)}_{k} =\sum_{i=1}^{\infty}w^{(t)}_{ik}\delta_{\theta_{i}} & \left(W^{(t)}_{1},\ldots,W^{(t)}_{p}\right) \sim \mbox{CRM}(\rho^{(t)},\lambda)\\ N^{(t)}_{k} =\sum_{i=1}^{\infty}\sum_{j=1}^{\infty}n^{(t)}_{ijk}\delta_{(\theta_{i},\theta_{j})} & N^{(t)}_{k}\mid W^{(t)}_{k}\sim\text{Poisson}\left(W^{(t)}_{k} \times W^{(t)}_{k}\right)
 \end{array} \end{aligned} \label{eq:Nhierarchy}%
\end{align}
where $n_{ijk}^{(t)}$ represent latent multiedges between nodes $i$ and $j$ within community $k$. 
The overall observed process is 
\begin{align}
{N}^{(t)}=\sum_{i,j} n^{(t)}_{ij}\delta_{\theta_i,\theta_j},
\end{align}
where ${n}^{(t)}_{ij}=\sum_{k=1}^p n_{ijk}^{(t)}$
is the total number of links between $i,j$ and ${n}^{(t)}_{ji}={n}^{(t)}_{ij}$ according to%(by summing over all communities)%, we get the total number of multi edges between two nodes:
\begin{align}
{n}^{(t)}_{ij}|(w^{(t)}_{\ell 1},\ldots,w^{(t)}_{\ell p})_{\ell=1,2,\ldots}  &  \sim\left\{
\begin{array}
[c]{ll}%
\text{Poisson}\left(2\sum_{k=1}^{p}w^{(t)}_{ik}w^{(t)}_{jk}\right) & i\neq j\\
\text{Poisson}\left(\sum_{k=1}^{p}(w^{(t)}_{ik})^2\right) & i=j
\end{array}
\right.
\label{eq:poi_nij}
\end{align}
which is a non-negative Poisson factorization~\citep{todeschini2020exchangeable, gopalan2015scalable, Ball2011, Psorakis2011}.

Although in this work we focus on multigraphs, note that if one is interested in the binary graph one could obtain it by setting a binary edge $z_{ij}^{(t)}=\mathbbm{1}_{n_{ij}^{(t)}>0}$ treating $n_{ij}^{(t)}$ as latent or, equivalently, using as link probability:
\begin{align}
{z}^{(t)}_{ij}|(w^{(t)}_{\ell 1},\ldots,w^{(t)}_{\ell p})_{\ell=1,2,\ldots}  &  \sim \left\{
\begin{array}
[c]{ll}%
\text{Bernoulli} (1-e^{-2\sum_{k=1}^{p}w^{(t)}_{ik}w^{(t)}_{jk}}) & i\neq j\\
\text{Bernoulli}(1-e^{-\sum_{k=1}^{p}(w^{(t)}_{ik})^2}) & i=j.
\end{array}
\right.
%\right\}
\label{eq:ber_zij}
\end{align}
%
%(With some abuse of notation) we will write the matrix of connections (multigraph) for a given time $t$ as ${N}^{(t)}$ so that its $(i,j)$th entry is given by ${n}^{(t)}_{ij}$.  
We construct the process so that marginally the vector of CRMs is a compound random measure~\citep{griffin2017compound}. The L\'evy measure is
\begin{align}\rho^{(t)}(dw)=\int w_{0}^{-p}F^{(t)}\left(\frac{dw^{(t)}_{1}}{w_0},\dots, \frac{dw^{(t)}_{p}}{w_0}\right) \rho_0(dw_0),
\label{eq:rho_t}
\end{align} 
resulting in weights factorized as $w^{(t)}_{ik}=w_{i0}\beta^{(t)}_{ki}$, where the time dependence comes through the evolution of the CRM %and not the base parameter $w_{i0}$
\begin{align}
W^{(t)}_k=\sum_{i=1}^\infty w_{0i}\beta^{(t)}_{ki}\delta_{\theta_i}.
\label{eq:W_t_k}
\end{align}
%
%In the dynamic community case, the sociability parameters need to
%be dependent over time, so that we may model the smooth evolution of the memberships of the nodes in time.
%We consider the sequence of random measures $$(W^{(t)})_{t=1,2,...}$ %to follow a Markov model, such that $W^{(t)}$ is marginally distributed as a Compound CRM with Gamma marginals for the scores, i.e. at each time $t$ the scores are distributed as \begin{align}\label{eq:GammaF}
Following~\cite{todeschini2020exchangeable} we draw the weights $w_{0i}$ choosing as L\'evy measure $\rho_0$ the one of a generalized gamma process GGP($\alpha,\sigma,\tau$)
\begin{equation}
\rho_0(dw_0)=\frac{1}{\Gamma(1-\sigma)}w_0^{-1-\sigma} e^{-w_0 \tau}dw_0
\label{eq:LevyGGP}
\end{equation}
for $\sigma \in (\infty, 1), \tau>0, \alpha>0$. The scores $(\beta_k^{(t)})_{k=1}^p$ are drawn from $p$ independent Gamma$(a_k, b_k)$ distributions for each timestep
\begin{align}
\label{eq:F_t}
F^{(t)}(d\beta^{(t)}_{1},\ldots,d\beta^{(t)}_{p}) 
= \prod_{k=1}^p F^{(t)}_k(d\beta^{(t)}_k) 
= \prod_{k=1}^p (\beta^{(t)}_{k})^{a_{k}-1}e^{-b_{k}\beta^{(t)}_{k}}\frac{(b_{k})^{a_{k}}}{\Gamma(a_{k})}d\beta^{(t)}_{k}.
\end{align}
%For $t=1,\dots, T-1$ we have 
%$$\Gamma^{(t)}_{k}=\sum %\gamma^{(t)}_{ki} %\delta_{\theta_{i}}$$ and
%
%In the rest we will dealing with Compound CRMs for which the CRM is the GGP where as the distribution of the scores $k$ is written by $F_k$ and we will write 
% In your document
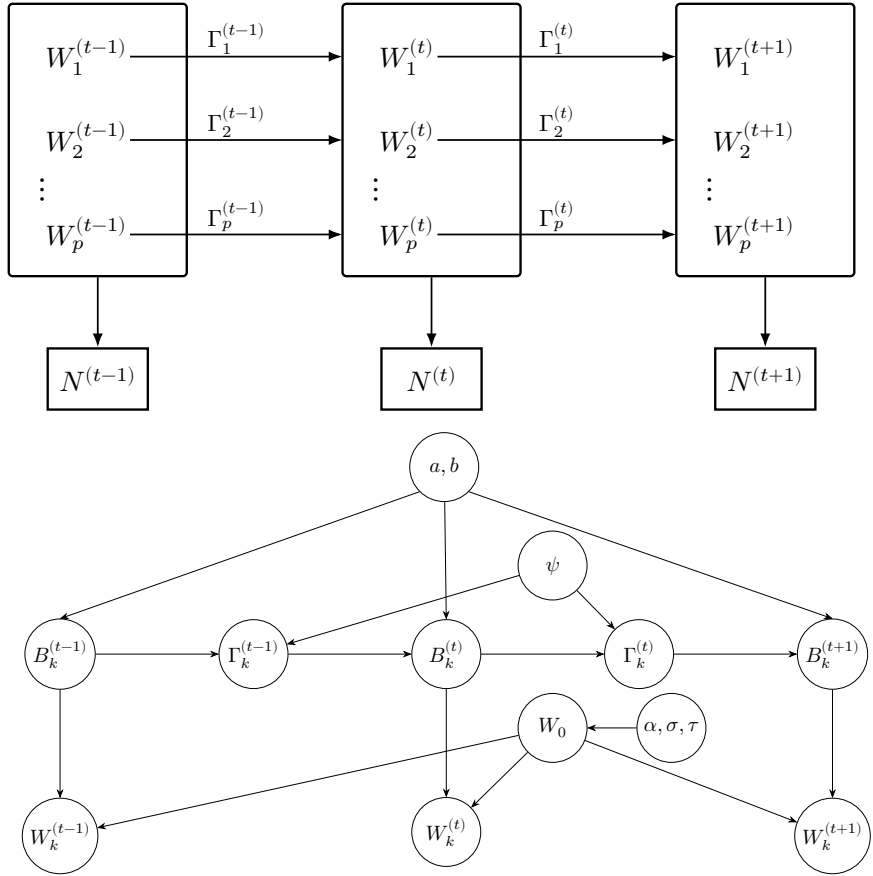
\begin{figure}[t!]
\centering
\centering
{\hspace{-0.15in}\resizebox{0.7\textwidth}{!}{\input{JASA_paper/latent_Markov_full}}}
\centering
{\hspace{-0.15in}\resizebox{0.7\textwidth}{!}{\input{JASA_paper/latent_Markov_k.tex}}}
\caption{Structure of the model. Top: high level view of the latent Markov structure for all $p$ measures. Bottom: Low level view of the latent structure in a given community $k$.} %Processes are defined in ~\ref{eq:processes},~\ref{eq:gammaprocess}
\label{fig:matrix}
\end{figure}
This construction gives us a time-varying compound generalized gamma process (CGGP):
%\begin{align}
%W^{(t)}_k=\sum_{i=1}^\infty w_0 \beta^{(t)}_{ki}\delta_{\theta_i}%\sim \text{ CGGP}(\alpha,\sigma,\tau, F^{(t)}_k)
%\end{align}
%
\begin{align}
W^{(t)}=(W^{(t)}_1,\dots,W^{(t)}_p) \sim \text{ CGGP}(\alpha,\sigma,\tau, F^{(t)})
\end{align} 
where $F^{(t)}$ is a product of Gamma distributions as shown in~\eqref{eq:F_t}.
\begin{figure}[t!]
    \centering
\includegraphics[width=12cm]{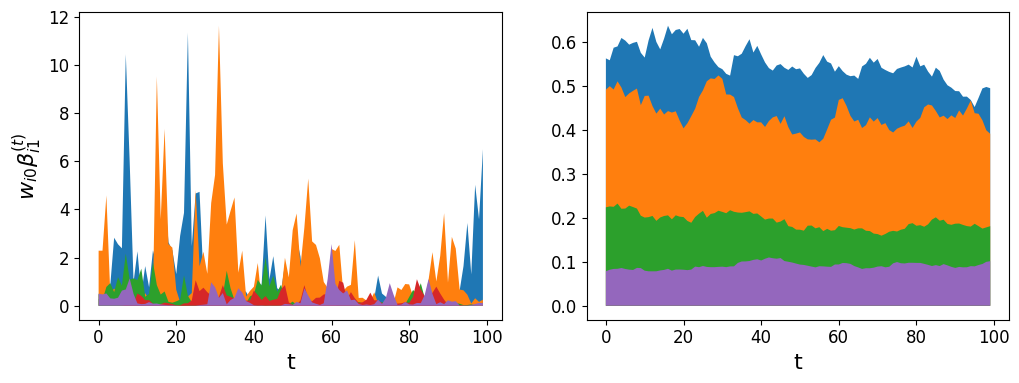}
    \caption{Temporal evolution of the weights of the first community of the $5$ highest degree nodes from graphs generated from a CGGP model with $\sigma=0.01, \tau=1, \alpha=2, p=2, T=100, a=b=(0.5,0.5)$, $\psi=2$ (left) and $\psi=2000$ (right).}
    \label{fig:weights_dependence_psi}
\end{figure}
We can write this process as a combination of $W_0=\sum_{i=1}^\infty w_{0i}\delta_{\theta_i}\sim\text{ CRM }(\rho_0, \lambda) $ and the $k-$th score process $ B_k^{(t)}=\sum_{i=1}^\infty \beta^{(t)}_{ki}\delta_{\theta_i}$. Their dependency will be explained below and they are shown graphically in Figure~\ref{fig:matrix}. Note that by normalizing the scores at each timestep we can visualize the proportions of affiliations of each node to each community as done in Figure~\ref{fig:toyexample}. 

\subsection{A Markov process for the dynamic memberships}
\label{sec:dependent_process}
To have temporal dependency and evolution of community affiliations, the scores evolve stochastically over time following the ideas of~\cite{Pitt2012}. The evolution is driven by augmentation variables $\gamma^{(t)}_{ki}$ giving rise to the $p$ latent processes $
\Gamma^{(t)}=(\Gamma^{(t)}_1,\dots,\Gamma^{(t)}_p)$ with $\Gamma^{(t)}_{k}=\sum_{i=1}^\infty \gamma^{(t)}_{ki} \delta_{\theta_i}$.
We construct them as below
\begin{align}
\label{eq:MarkovProcess}
\beta^{(t)}_{ki} \sim \text{Gamma} (a_k,b_k), \qquad 
\gamma^{(t)}_{ki}|\beta^{(t)}_{ki} &\sim \text{Gamma}(\psi,\beta^{(t)}_{ki})
\end{align}
and hence the derived conditional is 
\begin{align}
\label{eq:beta_posterior}
% \notag\\
     \beta^{(t)}_{ki} \mid \gamma^{(t)}_{ki} &\sim \text{Gamma}({a_{k}+\psi, \gamma^{(t)}_{ki}}+b_{k}).
\end{align}
%so that the scores are correlated across timesteps but marginally they are still Gamma distributed.
%where $\psi$
% tunes the correlation of the community memberships from one timestep to the next.
%
By taking the conditional law of $\beta^{(t+1)}_{ki}\;\mid\;\gamma^{(t)}_{ki}$ to be the same as that of $\beta^{(t)}_{ki}\;\mid\;\gamma^{(t)}_{ki}$,
%\begin{align}
%\label{eq:beta_conditional}
%\beta^{(t+1)}_{ki}\;\mid\;\gamma^{(t)}_{ki} \;\sim\; \mathrm{Gamma}(a_k+\psi,\, %\gamma^{(t)}_{ki}+b_k),
%\end{align}
the marginals for the scores at time $t$ and $t+1$ are the same $\mathrm{Gamma}(a_k,b_k)$.
%is a \emph{derived posterior distribution}, %obtained by Bayes' rule 
%from the joint model 
%\[
%\beta^{(t)}_{ki}\sim \mathrm{Gamma}(a_k,b_k), 
%\qquad 
%\beta^{(t+1)}_{ki}\sim \mathrm{Gamma}(a_k,b_k).
%\]
The above ensures that the processes $ B_k^{(t)}=\sum_{i=1}^\infty \beta^{(t)}_{ki}\delta_{\theta_i}$ and $W_k^{(t)}=\sum_{i=1}^\infty w^{(t)}_{ki}\delta_{\theta_i}$ have the same marginals, as explained in the next proposition. A high level graphical representation of the processes $W_k^{(t)}$ and $\Gamma_k^{(t)}$ is shown in Figure~\ref{fig:matrix} (top).% where a low level representation of what is happening for each community is show at the bottom, whereas a high level picture of all the processes taking place is shown on top.

%\gamma^{(t)}_{ki}\mid \beta^{(t)}_{ki}\sim \mathrm{Gamma}(\psi, \beta^{(t)}_{ki}).
%\]
%This construction is mirrored in the CRMs. Given $W^{(t)}\sim CGGP(\alpha,\sigma,\tau,F)$ and the auxiliary measure $\Gamma^{(t)}|W^{(t)}$ given by $\Gamma^{(t)}=(\Gamma^{(t)_1,\dots,\Gamma^{(t)}_p})$ with $\Gamma^{(t)}_k$

%The above construction ensures that \textcolor{red}{XXXX write from pitt and walker paper}. 
%We derive the expectation and variance of $\beta^{(t+1)}_{ki}\mid \beta^{(t)}_{ki}$ in the Appendix.
%
\begin{proposition}[Posterior of $W^{(t)}_k$ given $\Gamma^{(t)}$]
Consider $W^{(t)}$ the Compound GGP on $(0,\infty)\times \Theta$ with intensity $\nu^{(t)}(dw)=\rho^{(t)}(dw)\,\lambda(d\theta)$ where the L\'evy measure is defined in~\eqref{eq:rho_t}:
\begin{align}
W^{(t)}\sim \text{ CGGP}(\alpha,\sigma,\tau, F^{(t)}).
\end{align} 
For each time $t$ and community $k=1,\ldots,p$,
\[
W^{(t)}_{k} \;=\; \sum_{i=1}^\infty \beta^{(t)}_{ki}\,w_{i0}\,\delta_{\theta_i}
%\sim \text{CGGP}(\alpha,\sigma,\tau,F_k)
\]
with $w_{i0}$ from~\eqref{eq:LevyGGP} and $\beta^{(t)}_k,\gamma^{(t)}_{ki}\mid\beta^{(t)}_{ki}$ from~\eqref{eq:MarkovProcess}.
%\begin{align}
%\Gamma^{(t)}_{k}=\sum \gamma^{(t)}_{ki} \delta_{\theta_i}
%\qquad 
%\gamma^{(t)}_{ki}\mid \beta^{(t)}_{ki} \;\sim\; \mathrm{Gamma}(\psi, \beta^{(t)}_{ki}).
%\label{eq:gammaprocess}
%\end{align}
The posterior $W^{(t)}_{k}| \Gamma^{(t)}_{k}=\sum \gamma^{(t)}_{ki} \delta_{\theta_i}$ is 
\begin{equation}\label{eq:posterior_Wt_gammat}
    W^{(t)}_{k}\mid \Gamma^{(t)}_{k}=\sum_{i=1}^\infty \tilde\beta^{(t)}_{ki} w_{0i}\delta_{\theta_i}
\end{equation}
where
\begin{equation*}
\tilde{\beta}_{k}^{(t)}\sim \tilde F_k^{(t)}=\mathrm{Gamma}(a_k+\psi,b_k+\gamma^{(t)}_{k})
\end{equation*}
Therefore, the posterior is a compound GGP with score function $\tilde{F}^{(t)})$:
$$ W^{(t)}\mid \Gamma^{(t)}\sim \text{CGGP}(\alpha,\sigma,\tau,\tilde{F}^{(t)}).$$
\end{proposition}

If $W^{(t+1)}_{k}\mid \Gamma^{(t)}_{k}$ is distributed as Equation~\eqref{eq:posterior_Wt_gammat}, then $W^{(t+1)}_{k}$ and $W^{(t)}_{k}$ have the same marginals. %: $CGGP(\alpha,\sigma,\tau,F_k)$. 
Dependence between $W^{(t)}_k$, $\Gamma^{(t)}_k$, $B^{(t)}_k$, $W_0$, $N_k^{(t)}$ is shown in Figure~\ref{fig:matrix} (bottom).
\subsubsection{Interpretation of the parameters and hyperparameters}\label{subsec:interpretation}
The hyperparameters $\alpha, \sigma, \tau$ tune the overall properties of the graph (size, sparsity, degree distribution). %$\psi$ tunes the correlation of \textcolor{red}{the scores} between timesteps, 
The vectors $a=(a_1,\dots,a_p),b=(b_1,\dots b_p)$ tune the distribution of the scores $\beta^{(t)}_{ki}, k=1,\dots,p$, which represent the amount of affiliation of node $i$ to community $k$ at timestep $t$. $w_{i0}$ can be seen as base sociabilities of each node (indeed, the link function is increasing in $w_{i0}$). They act as degree correction parameters, taking care of sparsity and degree properties. Together with $(\beta^{(t)}_{ki})_k$ they define the sociability weights $w^{(t)}_{ik}$, which determine the number of links between nodes at time $t$.
This construction can clearly model mixed memberships as the vectors $\beta^{(t)}_{ki}$ can have more than one nonzero values. %The scores and thus the sociability parameters evolve in time giving dynamic memberships and dynamic connectivity. 
The community scores evolve through a Markov process with
%$\beta^{(t)}_{ki}$.
the distribution of $\beta^{(t+1)}_{ki}$ depending on $\beta^{(t)}_{ki}$ through $\gamma^{(t)}_{ki}$. The parameter $\psi$ governs the strength of correlation between timesteps, with larger $\psi$ implying higher correlation across time and hence smoother changes in time as shown in Figure \ref{fig:weights_dependence_psi}, where we generated graphs with the same hyperparameters except for $\psi$, which is $\psi=2$ (left) and $\psi=200$ (right).

\subsection{Sparsity and Power-Law Properties}\label{subsec:asymptotics}
%We can use Remark 5 in \cite{caron2023sparsity} to prove sparsity, power-law and clustering properties of our model, marginally over time. 
Here, we prove some asymptotic results. To do so, we phrase our proposal in the more general setting of \cite{caron2023sparsity}, which studied the asymptotic properties of graphs generated by the so-called graphex process. Since generally $W_k^{(t)}(\mathbb{R}_+)$ and $N^{(t)}(\mathbb{R}_+^2)$ will be infinite, they will showcase an unbounded number of active nodes and edges. To model finite graphs and define an asymptotic process, we restrict the point process on a portion of the real line identified by those nodes satisfying $\theta\le\alpha$ and discard nodes without connections, which in the case of infinite activity CRMs will be infinite even for the restricted process. The original, infinite dimensional graph, is obtained by the limit as $\alpha$ tends to infinity. 
We define the following summary statistics related to the restricted graphs (through the subscript $\alpha$):
\begin{align}
\label{eq:summary_stats}
    D_{\alpha i}^{(t)}:=\frac{1}{2}\sum_{k\ge 1}N^{(t)}_{ik}\mathbbm{1}_{\theta_k\le \alpha}, \hspace{0.5cm}
N_{\alpha}^{(t)}:=\sum_{i=1}^\infty\mathbbm{1}_{D_{\alpha i}^{(t)}\ge 1}\mathbbm{1}_{\theta_i\le \alpha}
\end{align}
which are respectively the number of connections of node $i$ at time $t$ and the number of nodes which display at least one connection at that timestep. We call these active nodes. Switching from the multigraph to the simple\footnote{A simple graph is a binary, undirected, graph with no self-loops.} graph process, we define $Z_{ij}^{(t)}=1$ if $N_{ij}^{(t)}>0$. Thus, we define 
\begin{align}
    N_{\alpha j}^{(t)}:=\sum_{i=1}^\infty\mathbbm{1}_{\sum_j Z_{ij}^{(t)}=j}\mathbbm{1}_{\theta_i\le \alpha}, 
    \qquad
    E_{\alpha}^{(t)}:=\frac{1}{2}\sum_{i\neq j}N_{ij}^{(t)}\mathbbm{1}_{\theta_i\le\alpha,\theta_j\le\alpha}
\end{align}
as the number of active nodes with degree $j$ as and the number of edges at time $t$ respectively. 
We recall that an increasing family of graphs with $N_\alpha^{(t)}$ active nodes and $E_\alpha^{(t)}$ edges is dense if 
$E_\alpha^{(t)}=\Theta ((N_\alpha^{(t)})^2)$
and sparse if $E_\alpha^{(t)}=o((N_\alpha^{(t)})^2)\text{ as }t$ grows to infinity. We can now prove the following propositions on sparsity and degree distribution.

\begin{proposition}[Sparsity]\label{prop:sparsity}
    Let $N_{\alpha}^{(t)}$ and $E_{\alpha}^{(t)}$ be respectively the number of active nodes and edges at time $t$. As $\alpha$ tends to infinity and for every $t$,
    \begin{align*}
    E_{\alpha}^{(t)}\asymp\begin{cases}
     (N_{\alpha}^{(t)})^{2/(1+\sigma)} \quad & \text{for }\sigma\in(0,1)\\(N_{\alpha}^{(t)})^2/\log(N_{\alpha}^{(t)})^2 \quad &\text{for }\sigma=0\\
     (N_{\alpha}^{(t)})^2\quad &\text{for }\sigma<0.
    \end{cases}
    \end{align*}
\end{proposition}
This result implies that for $\sigma<0$ the resulting asymptotic graph is dense, for $\sigma\in(0,1)$ it is sparse and for $\sigma=0$ a transition regime known as almost density. It is also worth noting that a distinction on the sparsity and density could be obtained also by applying Proposition 1 of \cite{todeschini2020exchangeable}, which proves that a finite activity CRM gives dense graphs, while an infinite activity CRM sparse graphs. Indeed, $\sigma\in(0,1)$ and $\sigma<0$ characterize respectively infinite and finite activity GGPs. Our proposition \ref{prop:sparsity} here is stronger since it characterizes exactly the rates of growth of the edges as a function of the number of nodes.

\begin{proposition}[Degree distribution]\label{prop:deg_distr}
Let $N^{(t)}_{\alpha}$ be the number of nodes at time t for the graph restricted at $\alpha$, and $N^{(t)}_{\alpha,j}$ be the number of nodes with degree $j\ge 1$ at time $t$. As $\alpha$ tends to infinity and $\sigma\in(0,1)$
\begin{align}\label{prop:asymp_degree_distr}
\frac{C_1}{C_2}\frac{\sigma \Gamma(j-\sigma)}{j!\Gamma(1-\sigma)}\leq \lim_{\alpha\rightarrow\infty}\frac{N_{\alpha j}^{(t)}}{N_{\alpha}^{(t)}}\leq  \frac{C_2}{C_1}\frac{\sigma \Gamma(j-\sigma)}{j!\Gamma(1-\sigma)},\quad\text{for }j\ge 1
\end{align}
where $C_1,C_2$ are positive real numbers. For $\sigma\le 0$, 
\begin{align*}
    \frac{N_{\alpha j}^{(t)}}{N_{\alpha}^{(t)}}\rightarrow  0,\quad\text{for }j\ge 1
\end{align*}
almost surely as $\alpha$ tends to infinity.
\end{proposition}

We can rewrite \eqref{prop:asymp_degree_distr} using the fact that for $j$ large $\frac{\sigma \Gamma(j-\sigma)}{j!\Gamma(1-\sigma)}\sim \frac{\sigma}{\Gamma(1-\sigma)}j^{-\sigma-1}$ and therefore for $\sigma\in(0,1)$ the asymptotic graph displays a power-law degree distribution with exponent $1+\sigma$ for large degrees. For the proofs of Propositions~\eqref{prop:sparsity} and \eqref{prop:deg_distr} see Section A1 %~\ref{sec:A} 
of the Supplementary.
\subsection{Simulation}
\label{sec:simulation}
%\textcolor{red}{As in Todeschini we can say something about simulation in an efficient way.+ complexity, generate w0, beta and gamma for all the timesteps and then we generate at teach timestep the graph by todeschini.}
%\textbf{Naive Simulation}\\
The hierarchical nature of the model in~\eqref{eq:Nhierarchy} suggests a (brute force) simulation as below. 
\begin{enumerate}
    \item Sample $(w_{i0},\theta)_{i=1,2,\dots}$ from a Poisson process with mean measure \\%GPP~\eqref{eq:LevyGGP}
$\rho_0(dw_{01},\dots,dw_{0p})\lambda(d\theta)1_{\theta \in [0,\alpha]}$ and construct $\beta^{(t)}, \gamma^{(t)}$ for $t=1,\dots,T$ from~\eqref{eq:MarkovProcess}.
    \item For each pair of nodes sample $n^{(t)}_{ij}$ from~\eqref{eq:poi_nij}.
\end{enumerate}

As explained in~\cite{todeschini2020exchangeable} and ~\cite{Caron2017graph} this has several challenges. In step $1.$ the measure is infinite and needs to be truncated at some threshold $\epsilon$ to generate a finite number of weights above $\epsilon$. The truncation affects the $w_0$s which would be sampled from the truncated measure $\rho_0^\epsilon$. Equation~\eqref{eq:rho_t} becomes
$\rho^{(t)}_\epsilon(dw^{(t)})=\int_\epsilon^\infty w_{0}^{-p}F^{(t)}\left(\frac{dw^{(t)}_{1}}{w_0},\dots, \frac{dw^{(t)}_{p}}{w_0}\right) \rho_0(dw_0)$
and $\int_{\mathbb{R}^p_+} \rho^{(t)}_\epsilon(dw_1^{(t)},\dots,dw_p^{(t)}) < \infty$, indicating that a smaller $\epsilon$ would lead to a better approximation. Secondly, in $2.$ we need to consider all pairs of nodes, which is $O(N^2)$ for $N$ nodes. We therefore expand on~\cite{todeschini2020exchangeable} and suggest the following, faster simulation scheme.\\
\newpage
\textbf{Simulation Algorithm}
\begin{enumerate}
    \item Sample the full set of parameters:
    \begin{enumerate}
        \item 
    Sample $(w_{i0},\theta_i)_{1\le i \le N}$ from a Poisson process with mean measure $\rho_0(dw_0)\lambda(d\theta)1_{\{w_0>\epsilon, \, \theta \in [0,\alpha]\}}$, giving rise to a finite number of nodes $L$. 
    \item For $t=1,\dots,T$ sample $(\beta^{(t)}_{ki})_{1\le i \le L, 1\le k \le p }$ through the auxiliary variables $(\gamma^{(t)}_{ik})_{1\le i \le L, 1\le k \le p }$ from~\eqref{eq:MarkovProcess} (note that $\gamma^{(t)}_{ik}$ go up to $T-1$ and not T).\\ 
    These give rise to $w^{(t)}_{ik}=w_{i0}\beta^{(t)}_{ki}$, the corresponding measure $W^{\epsilon(t)}_{k,\alpha}=\sum_{i=1}^L w^{(t)}_{ik}\delta_{\theta_i}$ and their total masses $W^{*\epsilon(t)}_{k,\alpha}=\sum_{i=1}^L w^{(t)}_{ik}$,  for $k=1,\dots,p$.
    \end{enumerate}
    \item For each $t=1,\dots, T$ sample a graph:%following~\cite{todeschini2020exchangeable}
    \begin{enumerate}
        \item For $k=1,\dots,p$ sample the total number of multiedges 
        \\$N^{*\epsilon(t)}_{k,\alpha}\mid W^{*\epsilon(t)}_{k,\alpha}\overset{\text{ind}}{\sim}\text{Poisson}\left((W^{*\epsilon(t)}_{k,\alpha})^2\right)$
        \item For $k=1,\dots,p$ and $\ell=1,\dots, N^{*\epsilon (t)}_{k,\alpha}$ sample the sender and receiver nodes $U^{(t)}_{k\ell j} \mid W^{\epsilon(t)}_{k,\alpha} \overset{\text{ind}}{\sim} W^{\epsilon(t)}_{k,\alpha}/W^{*\epsilon(t)}_{k,\alpha}$ for $j=1,2$.
        \item Obtain the multigraph $N^{\epsilon (t)}_{k,\alpha}=\sum_{\ell=1}^{N^{*\epsilon (t)}_{k,\alpha}} \delta_{U^{(t)}_{k\ell1},U^{(t)}_{k\ell2}}$ per community $k$ and sum over communities $N^{\epsilon (t)}_{\alpha}=\sum_{k=1}^p N^{\epsilon (t)}_{k, \alpha}$.
        \end{enumerate}
\end{enumerate}

%$w_{tik}$ is the sociability weight of node $i$ with respect to community $k$ at time $t$ which tunes the probability of connection between i and j at time t. It comprises of $w_0$ and $\beta_{tik}$ where and $\beta_{ik}$ is the affiliation of node $i$ to community $k$ at time t. 

%On identifiability ?? As noted both in ~\cite{Caron2017} and ~\cite{Todeschini2016}

%\begin{itemize}
    %\item maths of the model
    %\item interpretation: ie evolution of communities, evolution of memberships, mixed memberships
    %\item Focus on a toy example and add a graphical representation of the model.
%    \item Something on Parameter identifiability. either here or in experiments we need to talk about identifiability of params and also of the label switching problem and our solution.
    %\item A dependent Generalised Gamma Process for the affiliation
    %\item Varying connectivity parameters versus varying group membership
%parameters
%\end{itemize}
%
\section{Inference}
\label{sec:inference}
%\textcolor{red}{change $K$ to $K_\alpha$ or say that with some abuse of notation we denote below $K$ TO BE $K_\alpha$}?
%$N_{t,\alpha}
Assume that we observe multiple connections over time between a set of $N_{\alpha}$ nodes. Note that not all nodes appear in all timesteps, hence the observed numbers of nodes with at least one connection per timestep are denoted by the overlapping sets $N_{1,\alpha},\dots, N_{T,\alpha}$ %in other words some nodes might have zero connections at one timesteps and many connections at another.
and we denote by $N_{\alpha}$ the union of them. %$K=K_1\cup K_2\cup \dots K_T$. 
%all nodes that appear at at least one timestep. 
We denote our observed data as
$\mathcal{D}
=\{\left(n^{(1)}_{ij}\right)_{1\leq i,j\leq N_{\alpha}},\dots, \left(n^{(T)}_{ij}\right)_{1\leq i,j\leq N_{\alpha}}\},$
%$\left(n^{(t)}_{ij}\right)_{1\leq i,j\leq N,\text{ and } t=1,\dots,T}$ 
where at each timestep, some nodes might be of zero degree as explained above. % of the $N$ nodes have at least one connection.
Our goal is to infer the set of positive parameters $
(w^{(t)}_{i1},\dots,w^{(t)}_{ip})_{i=1,\dots,N_{\alpha}}$ for $t=1,\dots,T$. These depend on the hyperparameters $\xi=\left(\alpha,\sigma,\tau,\psi,a=(a_1,\dots,a_p),b=(b_1,\dots,b_p)\right)$ of the model and the transition variables $\left(\gamma^{(t)}_{i1},\dots,\gamma^{(t)}_{ip}\right)_{i=1,\dots,N_{\alpha}}$. 
Note that we will not attempt to estimate the node locations $(\theta_i)_i$ as they are not likelihood identifiable. If one considers the model using the infinite activity GGP, then they should also estimate the positive parameters that are associated with nodes with no connections (not present in the set of $N_{\alpha}$). These are only identifiable through their sum which we can be denoted as $(\tilde W^{(t)}_1,\dots, \tilde W^{(t)}_p)$. Therefore the full posterior would be
$$p\left( (\xi,w^{(t)}_{i1},\dots,w^{(t)}_{ip},\gamma^{(t)}_{i1},\dots,\gamma^{(t)}_{ip},W^{(t)}_1,\dots, \tilde W^{(t)}_p)_{i=1,\dots N_{\alpha},t=1,\dots,T}|{n}^{(1)},\dots,{n}^{(T)}\right),$$ where ${n}^{(t)}=(n^{(t)}_{ij})_{1\leq i,j\leq N_{\alpha}}.$
Various approaches can be taken for inference, such as maximum likelihood estimation or Bayesian inference either with MCMC sampling or a variational approximation. An algorithm for Bayesian inference with MCMC sampling can be devised by adapting the ideas of~\cite{todeschini2020exchangeable} to the dynamic case. However exact Bayesian inference under the infinite-activity GGP prior is computationally intensive.  here instead we rely on the proposal by~\cite{lee2023unified} to use as finite-dimensional independent and identically distributed approximation the exponentially tilted BFRY distribution (etBFRY)
\begin{align}
\label{eq:BFRY}
    w_{0i} & \overset{\text{iid}}{\sim} \text{etBFRY}\left(\alpha/L,\tau,\sigma\right), \text{ for } i=1,\dots,L
\end{align}
with density
\begin{align}\label{eq:etBFRY_density}
g_{\alpha,L,\tau,\sigma}(w) =\frac{\sigma w^{-1-\sigma}e^{-\tau w}\left( 1-e^{- (\sigma L /\alpha)^{1/\sigma}w} \right)  }{\Gamma(1-\sigma)\left\{ \left( \tau +(\sigma L /\alpha)^{1/\sigma} \right)^{\sigma} -\tau^{\sigma}\right\} }.
\end{align}
%Note that the prior used for ${w_0}$ is not the one originally used in \cite{Caron2017graph}, and subsequently in ~\cite{Todeschini2016}, who instead used a generalised gamma process $GG(\alpha,\sigma,\tau)$. 
As proved in \cite{lee2023unified}, the etBFRY distribution is guaranteed to converge to a GGP as the truncation level $L$ approaches infinity, etBFRY$(\alpha/L,\tau,\sigma)\overset{d}{\rightarrow}\text{GGP}(\alpha,\sigma,\tau)$, while preserving key structural properties such as sparsity and power-law degree behavior. (Note that the truncation level $L$ has to be bigger than the total number of active nodes $N_\alpha$.)

\textbf{Approximate Posterior.} 
%Denoting $\boldsymbol{N}^{(t)}=(n^{(t)}_{ij})_{1\leq i,j\leq N}$
The desired approximate posterior is
$$p\left( \xi,(w^{(t)}_{i1},\dots,w^{(t)}_{ip},\gamma^{(t)}_{i1},\dots,\gamma^{(t)}_{ip})_{i=1\dots L,t=1\dots T}|{N}^{(1)},\dots,{N}^{(T)}\right).$$
Considering a certain timestep $t$ we can write
\begin{align*}
    &p((w_{0i} ,\beta^{(t)}_{1i},\dots,\beta^{(t)}_{pi}, \gamma^{(t)}_{1i},\dots, \gamma^{(t)}_{pi}, \gamma^{(t-1)}_{1i},\dots, \gamma^{(t-1)}_{pi})_{i=1}^{L}, \xi \mid (n^{(t)}_{ij})_{i,j=1}^L) \notag\\
    &\propto p((n^{(t)}_{ij})_{i,j=1}^L|(w_{0i} ,\beta^{(t)}_{1i},\dots,\beta^{(t)}_{pi})_{i=1}^{L})p((w_{0i})_{i=1}^{L} )p((\beta^{(t)}_{1i},\dots,\beta^{(t)}_{pi})_{i=1}^{L}|(\gamma^{(t-1)}_{1i},\dots, \gamma^{(t-1)}_{pi})_{i=1}^{L})p(\xi)
\end{align*}
Using Bayes rule and~\eqref{eq:MarkovProcess},~\eqref{eq:beta_posterior} %~\eqref{eq:beta_conditional} 
we can compute 
$p((\beta^{(t)}_{ki})_{i=1}^{L}|(\gamma^{(t-1)}_{1i},\dots, \gamma^{(t-1)}_{pi})_{i=1}^{L})\propto p\left(\gamma^{(t)}_{ki}|\beta^{(t)}_{ki}\right)p\left(\beta^{(t)}_{ki}|\gamma^{(t-1)}_{ki}\right)$ which is recognized as a Gamma random variable
\begin{align}
\label{eq:beta_doubleconditional}
\beta^{(t)}_{ki}|\gamma^{(t-1)}_{ki}, \gamma^{(t)}_{ki}
\sim \text{Gamma}\left(a_{k}+2\psi, \gamma^{(t)}_{ki}+\gamma^{(t-1)}_{ki}+b_{k}\right).
\end{align} 

Note that~\eqref{eq:beta_doubleconditional} is only true for $t>1$, since $\gamma^{(0)}_{ki}$ is not defined and for $t=1$
$p(\beta^{(1)}_{ik}|\gamma^{(0)}_{ki}, \gamma^{(1)}_{ki})\sim p(\beta^{(1)}_{ik}|\gamma^{(1)}_{ki})$ is given by Equation~\eqref{eq:beta_posterior}. 
Therefore, the full posterior at $t$ is
\begin{align}\label{eq:approx_posterior}
    \log p((w_{0i} ,&\beta^{(t)}_{1i},\dots,\beta^{(t)}_{pi}, \gamma^{(t)}_{1i},\dots, \gamma^{(t)}_{pi}, \gamma^{(t-1)}_{1i},\dots, \gamma^{(t-1)}_{pi})_{i=1}^{L}, \xi \mid (n^{(t)}_{ij})_{i,j=1}^L) \notag\\
    \propto & \log p((n^{(t)}_{ij})_{i,j=1}^L|(w_{0i} ,\beta^{(t)}_{1i},\dots,\beta^{(t)}_{pi})_{i=1}^L) + \log p((w_0)_{i=1}^L) \\%+ \log p(\mathbf u|\mathbf w_0) 
    &+ \sum_{k}\log p(\beta^{(t)}_{ki})_{i=1}^L|(\gamma^{(t)}_{ki}, \gamma^{(t-1)}_{ki})_{i=1}^L)+ \log p(\xi)\notag\\ 
\propto & \left[\sum_{i=1}^L \sum_{j=1}^L \log\text{Poisson}\left({n^{(t)}_{ij};w_{0i}w_{j0}\sum_k \beta^{(t)}_{ki} \beta^{(t)}_{jk}}\right)\right]\notag\\
& + \left[\sum_{i=1}^L \left(\log\text{BFRY}(w_{0i}; \alpha/L,\tau,\sigma) 
%+ \log\text{tExp}(u_{i};w_{0i},t_{\alpha,\sigma}) 
+ \sum_k\log p\left(\beta^{(t)}_{ki}|\gamma^{(t-1)}_{ki}, \gamma^{(t)}_{ki}\right)\right) \right]+ \log p(\xi).\notag
\end{align}
A more detailed version of the posterior is found in Section A2 %~\ref{sec:A2} 
of the Supplementary.
%
%s\textcolor{red}{please check the posterior and L}
%
We perform Bayesian inference to approximate the posterior distribution in~\eqref{eq:approx_posterior} using gradient-based MCMC. %An inference scheme can be devised extending ideas from \cite{todeschini2020exchangeable} and \cite{naik2022bayesian} and considering HMC updates for the weights. 
Specifically we use Hamiltonian Monte Carlo via the NUTS~\citep{hoffman2014nuts} algorithm (within the probabilistic language NumPyro~\citep{phan2019composable}) which is well suited to the high-dimensional continuous parameter space induced by the dynamic latent variables, allowing efficient exploration of the posterior distribution.

As with other SBMs, the likelihood is invariant to permutations of the community labels, implying label switching. The temporal dependence induced by the latent Markov process encourages persistence of community structure across timesteps but does not eliminate label switching in posterior inference. Thus, caution is needed to interpret the community labels.%Node locations $(\theta_i)_i$ are not likelihood identifiable and are therefore not inferred, while remaining scale ambiguities between sociability weights and community scores are resolved through the model construction and prior specification.
\section{Experiments}
\label{sec:Experiments}
%In this section, we implement our model on synthetic and real-world networks. %\textcolor{red}{The code used for all experiments is available at xxxx.}
%
%
\begin{figure}[t!]
    \centering
    %\subfigure[t=1]
    {\includegraphics[width=5cm]{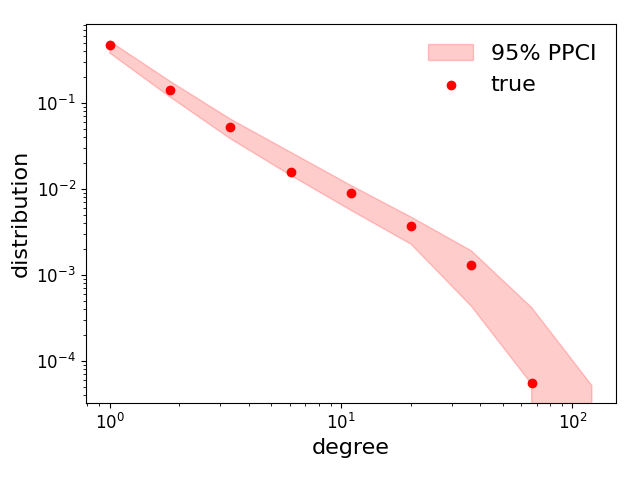}}
    %\subfigure[t=2]
    {\includegraphics[width=5cm]{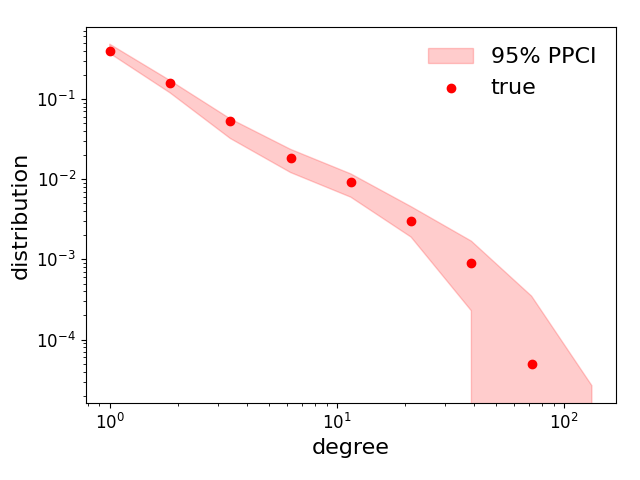}}
    %\subfigure[t=3]
    {\includegraphics[width=5cm]{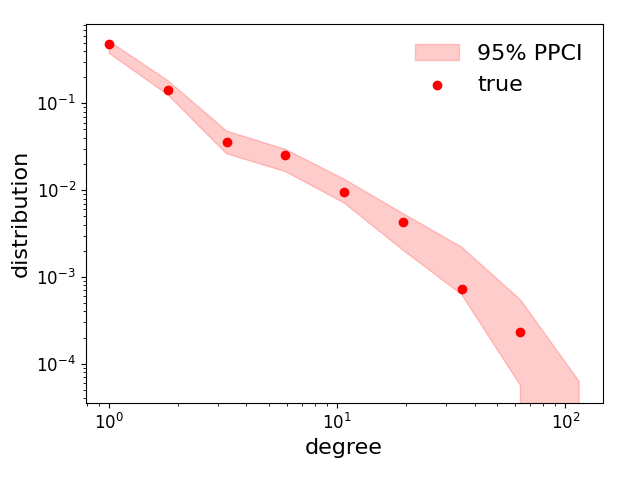}}
    \caption{Degree distribution in log-log scale for the synthetic dataset in $t=1$ (left), $t=2$ (middle), $t=3$ (right). Empirical degree in red dots, and $95\%$ PPCI in the shaded region.}
\label{fig:degree_distribution_synthetic}
\end{figure}
\subsection{Synthetic Experiment - validation of the approximation}
\begin{figure}[t!]
    \centering
    %\subfigure[score affiliations for community 1]
    {\includegraphics[width=6cm]{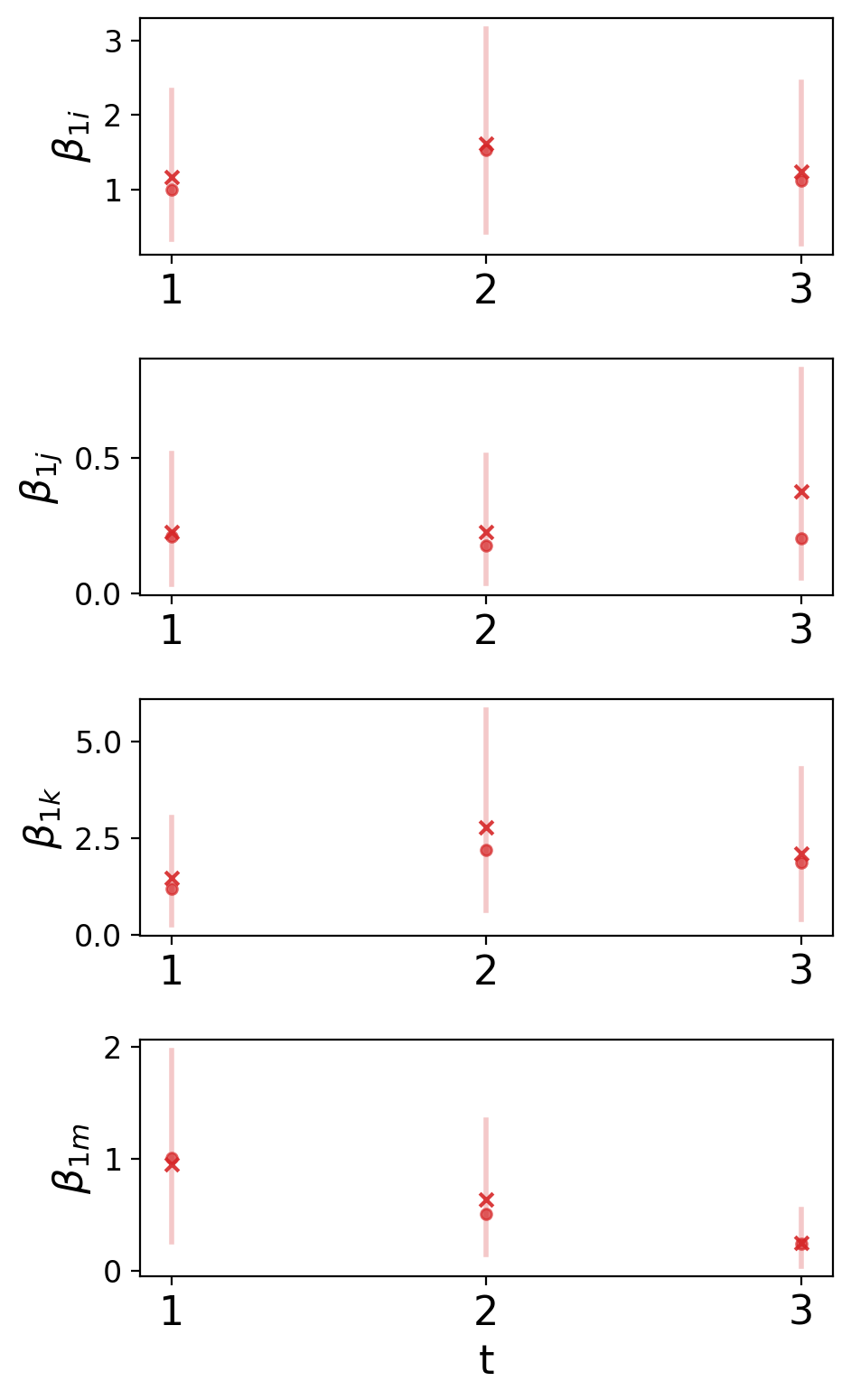}}
    %\subfigure[score affiliations for community 2]
    {\includegraphics[width=6cm]{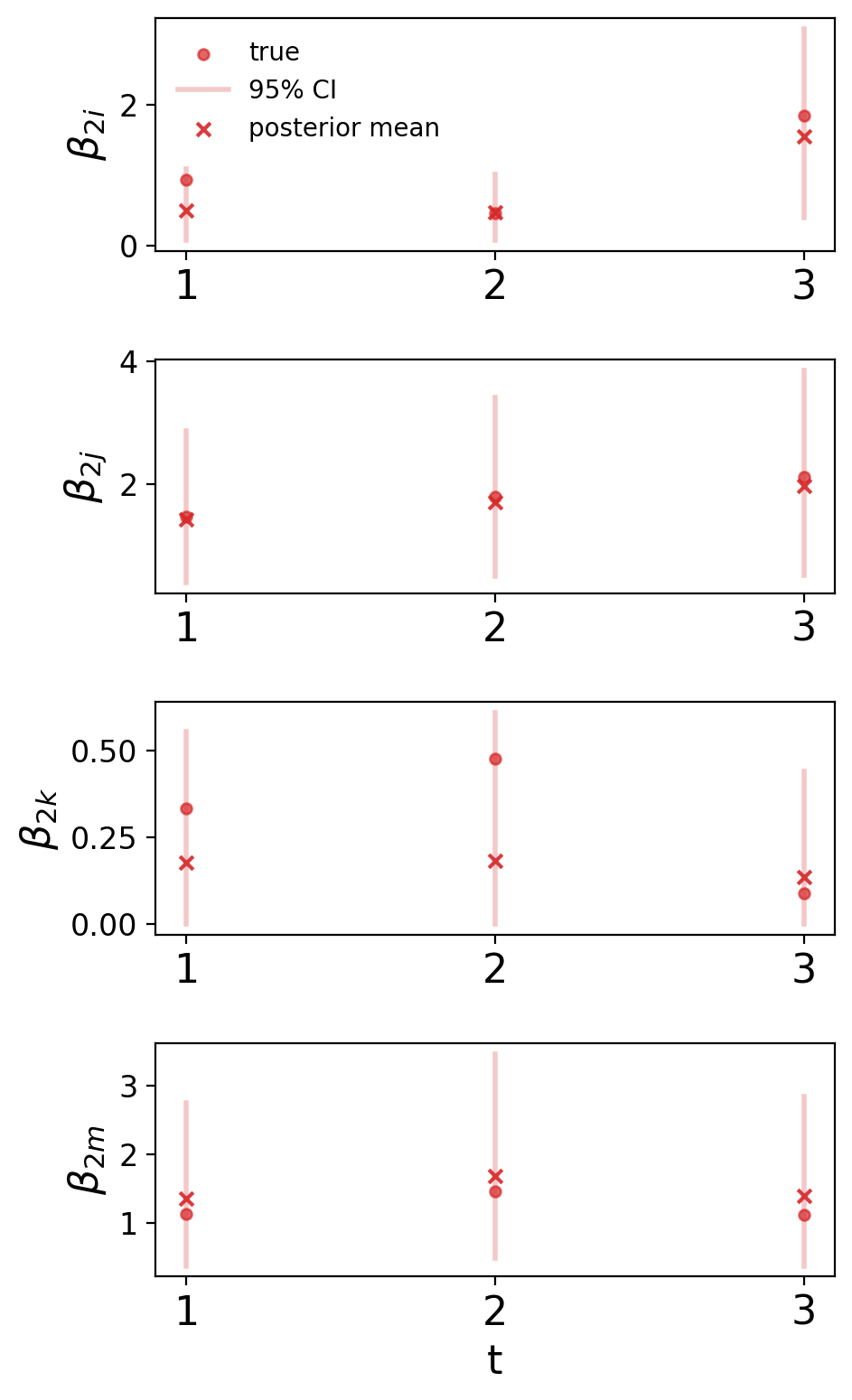}}
    \caption{ Temporal evolution of the scores for $t=1,2,3$ for four high degree nodes (named $i,j,k,m$) of the network simulated from the CGGP model. True scores are in dots, and 95\% CI in the vertical bars for community 1 (left) and community 2 (right).}
\label{fig:scores_high_degree_nodes_synthetic}
\end{figure}

In this section we first aim to validate our proposed approximation inference approach. We study the performance of the approximate inference on a synthetic dataset generated from the infinite Bayesian nonparametric model (see Section~\ref{sec:simulation}). %Results show good model fit and parameter recovery.
We generate an undirected multigraph for $T=3, p=2$ with parameters $\alpha =60, \sigma=0.2, \tau=1, \psi=5, a=(1,1), b=(1,2)$. To generate the graph we used a threshold $\epsilon=0.0001$. The synthetic graph has $1258$ nodes of which the active nodes are $327$ at $t=1$, $330$ at $t=2$ and $321$ at $t=3$. We set $L=1258$ and assume a positive Halfnormal prior on all unknown hyperparameters $\xi$. %We initialize from the static model?? 
To assess convergence we inspect the trace plots of the two MCMC chains and give the plots of the hyperparameters $\alpha, \sigma, \psi, a, b$ in %Figure~\ref{fig:trace_plots_synthetic} in
Section B1 %~\ref{sec:C1} 
of the Supplementary. %Similarly, in the Supplementary we show the log-posterior of the two chains in Figure~\ref{fig:log_posterior_synthetic} which suggests that they reach convergence. 
Regarding the parameters, namely the sociability weights $w_{0i}$ and scores $\beta^{(t)}_{ki}$ are well recovered. Specifically, we obtained $96\%$ coverage in the sociability weights and $95\%$ in the scores (on average from the two chains). 
Additionally, we show the scores for a set of high degree nodes. In Figure~\ref{fig:scores_high_degree_nodes_synthetic} we show their true value (red dots) and 95\% credible intervals (CI) in the shaded vertical bars 
for community 1 (left) and 2 (right). 
As the plot suggests, the algorithm recovers well the score values for each community. To further assess model fit, we generate $500$ graphs
from the posterior predictive distribution. % of the graph using the MCMC estimates of the hyperparameters, then generating the parameters and finally obtaining the predictive graphs. %We thus assess the performance of our model regarding
%degree distribution fit. 
As shown in Figure~\ref{fig:degree_distribution_synthetic}, the shaded region, which is the 95\% posterior predictive credible interval (PPCI) of the degree distribution, covers well the true empirical degree distribution. All together, these results mean that our model can capture well the power-law nature of such sparse graphs and in general recover the ground truth even when using an approximated inference.
%\textcolor{red}{We further assess the clustering performance of our model based on the adjusted Rand index (ARI). This index asses the agreement between the estimated and true latent structure for hard clustering. There is global and averaged ARI. Good values are close to 1. It is more difficult to recover the global one than the averaged over timesteps. Also how does the performance of ARI change wrt to more timesteps, or more separated communities?}
%
%
%
%
\subsection{Reuters Terror Dataset}
\textbf{Data Description.} We now run our model to learn the dynamic evolution of the latent communities of a real-world network, from the Reuters terrorist dataset\footnote{\url{https://sparse.tamu.edu/Pajek}}. This is a data set from the Reuters news agency concerning the 09/11/01 attack in the US. It is based on all stories published from 09/11/01 to 11/15/01, aggregated and shortened for our purposes. We aggregate the data from days into weeks and use the first $T=6$ weeks after the attack\footnote{Reuters  data (with preprocessing) and code for model implementation and experiments can be found in our anonymized repository \url{https://anonymous.4open.science/r/dynSNetOC-B16C/README.md}.}. Nodes are words and edges represent co-occurrence of words in a sentence, creating a undirected multigraph.
Not all words are present at each week, and only around $50\%$ of the nodes are active at each timesteps.
\begin{figure}[t!]
    \centering
    %\subfigure[t=1]
    {\includegraphics[width=5cm]{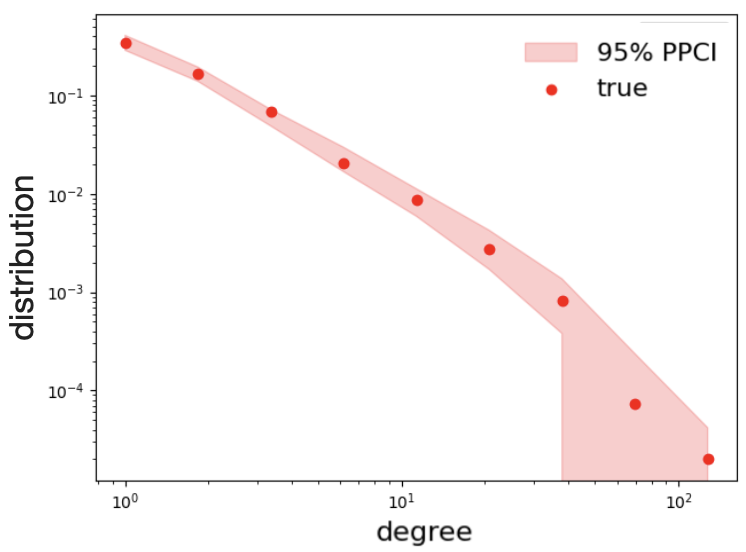}}
    %\subfigure[t=2]
    {\includegraphics[width=5cm]{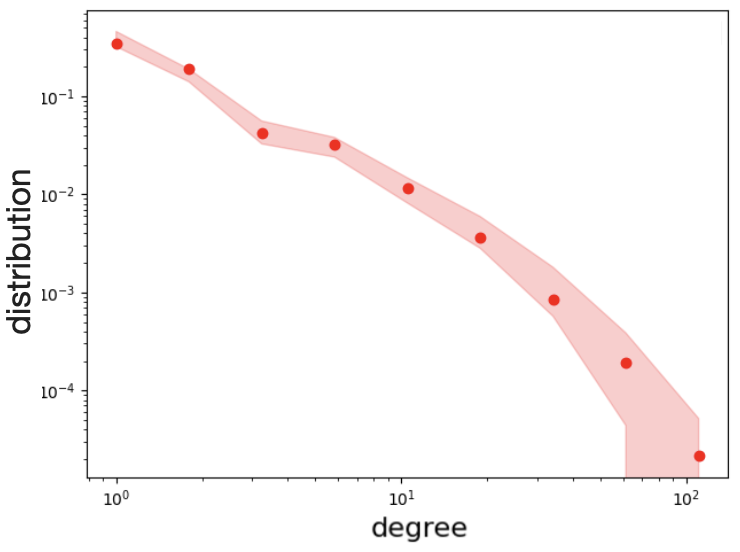}}
    %\subfigure[t=3]
    {\includegraphics[width=5cm]{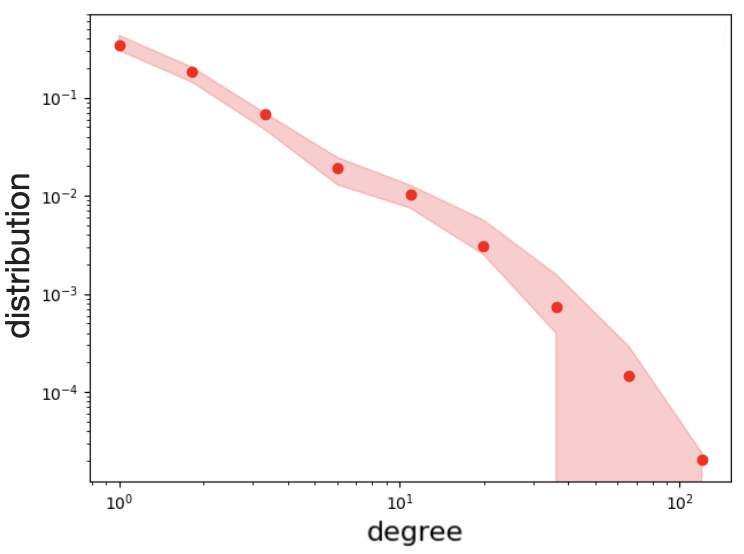}}
    %\subfigure[t=4]
    {\includegraphics[width=5cm]{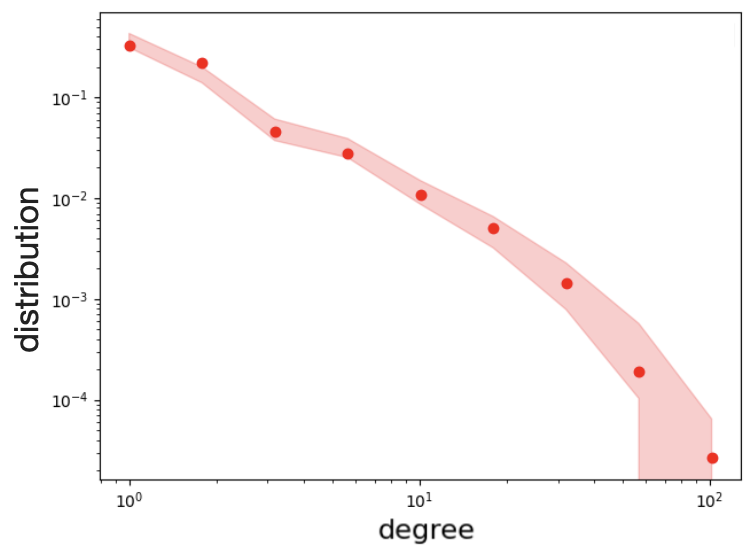}}
    %\subfigure[t=5]
    {\includegraphics[width=5cm]{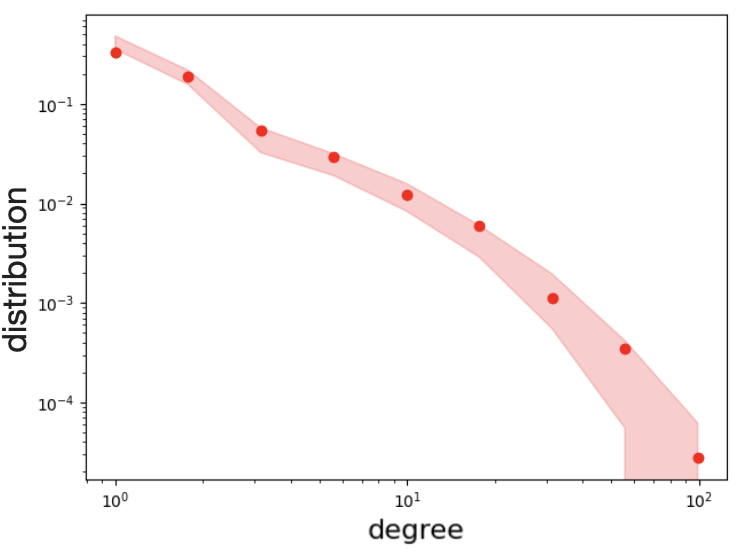}}
    %\subfigure[t=6]
    {\includegraphics[width=5cm]{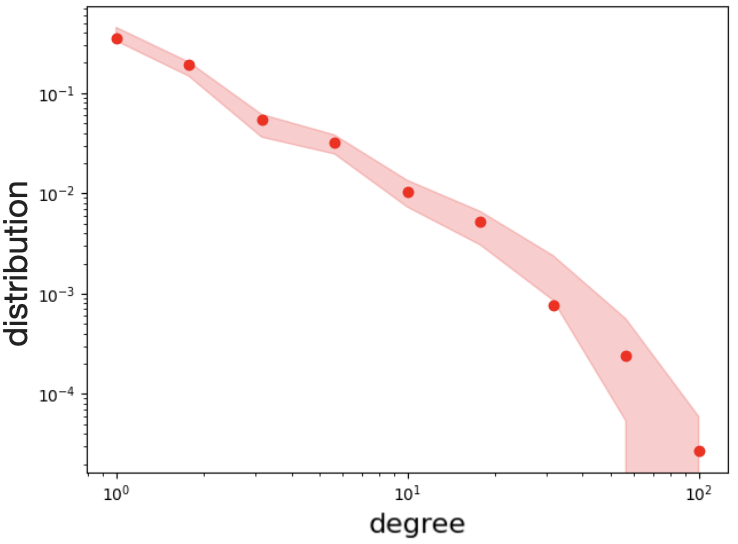}}
    \caption{Degree distribution in log-log scale for reuters for $t=1,2,3$ (top row) and $t=4,5,6$ (bottom row). Empirical degree in red dots, and $95\%$ PPCI in the shaded region.}
\label{fig:degree_distribution_reuters}
\end{figure}
\begin{table}[t!]
\caption{A high level view of representative words for each community for the 6 weeks.}
\label{tab:table_words_reuters}\centering
\begin{tabular}{|c|c|}
\hline
{Community} & {Representative words} \\ 
\hline
`Afhanistan’ &  Taliban, Al-Qaeda, Bin laden, Afghanistan\\ 
\hline
`anthrax’ &  office, post, security, senate, senator, newspaper, sample, Brokaw\\ 
\hline
`attack’ &  pentagon, new york, plane, hijack, WTC, attack\\ 
\hline
`political’ &  foreign minister, official, military, economy, financial\\ 
\hline
`security’ &  passenger, federal, hijacker, airline,
agent, FBI, man\\ 
\hline
\end{tabular}
\end{table}
Specifically, the numbers of active nodes are $(468, 525, 501, 474, 471, 525)$. Empirical sparsity\footnote{This is measured as number of edges divided by the square of number of active nodes.} at each timestep is $(0.02, 0.01, 0.01, 0.02, 0.02, 0.01)$, which indicates that the graphs are sparse. Our data consists of $6$ sparse graphs of $L=1200$ nodes where connections in the graph change in every timestep. The empirical degree distribution in each time is power-law, giving linear log-log degree distribution plots as shown by the red dots in Figure~\ref{fig:degree_distribution_reuters}.
\\
\textbf{Inference.}
 %an exponentially tilted BFRY on $w_0$.
 We run the NUTS algorithm on the data in order to estimate the unknown parameters and hyperparameters. % $\alpha, \sigma, \psi, a, b$.
%We initialize the estimation from the MAP estimates of the hyperparameters $\alpha, \sigma, \tau, a, b$ obtained from running the static model~\citep{todeschini2020exchangeable}. Subsequently, using this initialization for $\alpha, \sigma, \tau, a, b$ and a random initialization for $\psi$ we run the NUTS algorithm and 
We assume a vague positive Halfnormal prior on the unknown hyperparameters, but for $\tau$ we fix its value to avoid identifiability issues following~\cite{Caron2017graph,todeschini2020exchangeable}. We run one MCMC chain of $250,000$ iterations of which $130,000$ are discarded as burn-in. We give the trace plots of the identifiable parameters %which are $\log(\tilde \alpha),\psi, \sigma, a, \tilde b$ where $\tilde a =\alpha \sigma^\tau$ and $\tilde b=b\tau$ 
and the log-posterior (up to a constant) %in Figure~\ref{fig:trace_plots_reuters} 
in Section B2 %~\ref{sec:B2}
of the Supplementary. The trace plots show good convergence of the algorithm. 
Based on previous work and the nature of the dataset, we expect the communities to represent topics or themes that appeared in the news following the 9-11 attack.

%We set the number of communities per time step to $p=4$, due to the following reasons. 
Previous works~\citep{li2009discovering,DooleyCorman2002} and \cite{li2009discovering} study in different ways the communities throughout the weeks following the attack. Following their conclusions, we set $p=4$ in our model%expect to have $4$ communities at each timestep, which suggests to run the model for $p=4$.
\footnote{We also ran our model for $p=5$, but we concluded that $p=4$ was of better fit and better interpretation.}.
\begin{figure}[t!]
    \centering
    %\begin{minipage}{.45\textwidth}
\includegraphics[width=12cm]{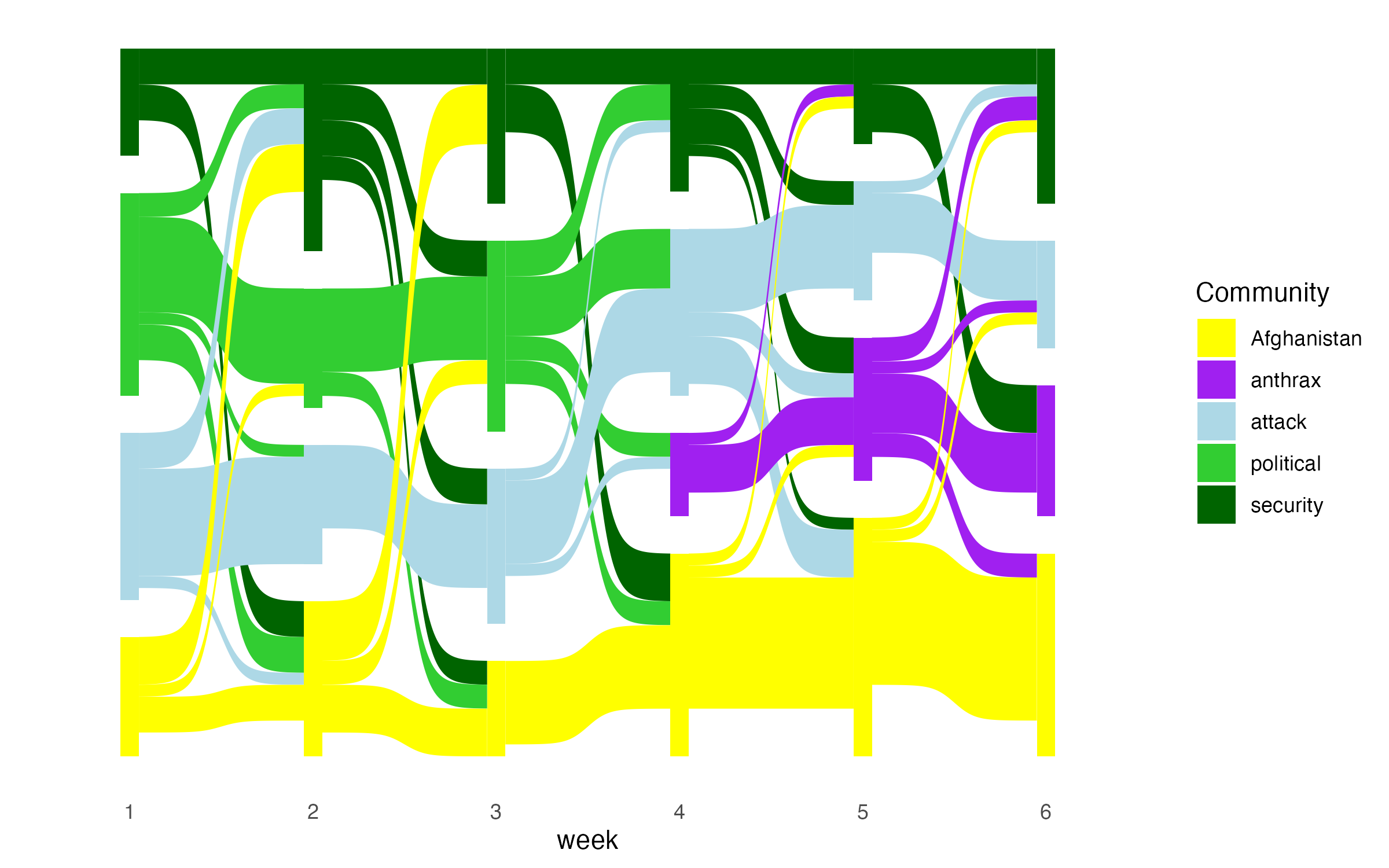}
    %\end{minipage}
    %\begin{minipage}{.45\textwidth}
    %\includegraphics[width=9cm, height=7cm]{figures/reuters_p5_T6_community_flow_top50degpertimestep.pdf}
    %\end{minipage}
%\includegraphics[width=8cm]{figures/reuters_p4_T6_community_flow_top50degpertimestep_DynSBM.pdf}
    \caption{Sankey plot on the dynamic behavior of the words memberships for reuters using our model. Colors correspond to communities which evolve along the weeks. %Words might switch community, or stop existing or new words might appear at later weeks. %We have used principal words (high degree words) as representatives of these communities. The thickness is proportional to the number of words.
} 
\label{fig:sankey_plots_reuters}
\end{figure}
Setting $p=4$ means that at any given timestep we have $4$ communities, but %since a community is just a container for a topic, 
the meaning of a community is allowed to change over time, with themes appearing and disappearing. Therefore, the total number of themes might be bigger than $p$. 
As we will show below, these topics evolve according to the historic events following the $9-11$ attack, such as threat of war, political moves or health hazards. Cross referencing our results with the events that happened those days can verify the correctness of our approach.
%
%We perform inference under the iid approximation to estimate the unknown parameters and hyperparameters. Note that we do not estimate $\tau$ to avoid identifiability issues. We set $tau$ to the value we obtained when we ran the static model with overlapping communities (\cite{todeschini2020exchangeable}) and then fix $\tau$ to that value. %\alpha, \sigma, a, b, \psi, (w_{0i})_i, (\beta^{(t)}_{ki})_{ki}$. For the interpretation of these parameters, see section~\ref{sec:model}. 

From the MCMC inference we get estimates of all the parameters and hyperparameters. To assess the model fit, the posterior predictive degree distribution is shown in Figure~\ref{fig:degree_distribution_reuters} which shows that the $95\%$ PPCI includes the empirical degree distribution.%

\textbf{Dynamic community interpretation.} We estimate the community affiliations for all the words which we analyze and give below a detailed description of the discovered communities and their evolution.
%\textbf{Dynamic evolution of communities.}
The communities we discover are `9-11 attack’, `Afghanistan’, `security’, `political response’, `anthrax’. As explained above, not all of them appear in all timesteps and words see their affiliation to the various communities varying with time. 
%The community anthrax only appears at week 4 onwards and is reflects a new emerging theme which was the anthrax letters that were mailed to media officers and political figures at the time. As suggested by the model and also demonstrated later on in Figure~\ref{fig:pie_plots_reuters}, the words have mixed memberships. For example the word attack is found to have high affiliations in both '9-11 attack and 'security communities. Additionally these memberships evolve in time. It is possible that a word might first belong primarily to a certain community but then switch. Then, if one wants to consider hard clustering, this word would switch cluster. 
%We explain below in detail the evolution and interactions of the communities, 
In Table~\ref{tab:table_words_reuters} we give a high level summary of the $5$ topics (communities). Words at the Table consider an overall picture of the communities from all $6$ weeks.%

We demonstrate the evolution of the communities along $6$ weeks in Figure \ref{fig:sankey_plots_reuters}. The Sankey diagram shows the evolution of the dynamic behavior of the communities across the timesteps. The plot shows the affiliations of the $50$ highest degree nodes for each week, assigning them to a community according to their highest affiliation. The thickness of the flow lines in the plot is proportional to the number of words in that community. 
\begin{figure}[t!]
    \centering
    %\begin{minipage}{.45\textwidth}
    \includegraphics[height=16cm]{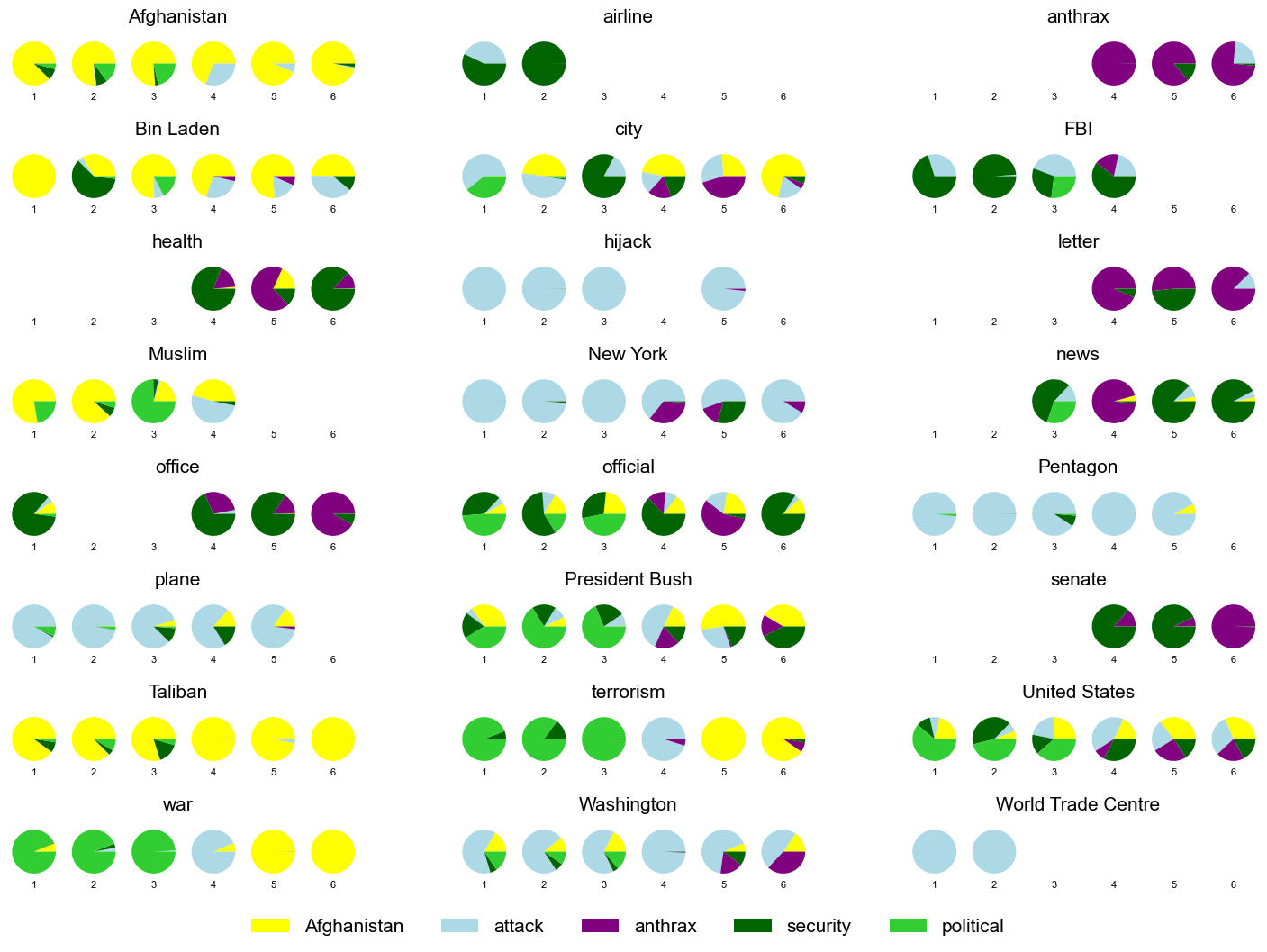}
    %\end{minipage}
    %\begin{minipage}{.45\textwidth}
    %\includegraphics[width=7cm]{figures/scores_p5_pie_plot_listwords.png}
    %\end{minipage}
    %\caption{Pie plots of weekly affiliation of principal words. On the left, for $p=4$, on the right $p=5$.}
    \caption{Pie charts with the \% of community affiliations for some high degree words at each timestep using our model. If a word is not present at a certain timestep (i.e. it is not in the 50 highest degree nodes) then its pie plot doesn't exist.}
\label{fig:pie_plots_reuters}
\end{figure}

Since the highest degree words are not necessarily the same across timesteps, some of them do not flow to subsequent weeks or might appear later in the graph (represented by the vertical bars that do not persist across timesteps). In Figure \ref{fig:pie_plots_reuters} we show the pie plots of the dynamic memberships from which one can see the evolution of the communities through their principal words. Based on this, we will now describe the communities.

%\indent $\quad \bullet$ 
Community `attack' describes the event of the attack which took place on $11$th Sept $2001$. Most of its words are associated with the WTC attack (e.g. \textit{World Trade Centre}, \textit{Pentagon}, \textit{plane}, \textit{hijacked}). On October $7$ (week $4$), the US invaded Afghanistan initiating what was known then as ``war on terror''. Hence, some of the words that belonged to the community `political' (\textit{Bush}, \textit{war}, \textit{terrorism}, \textit{US}...) on that week migrated towards `attack' changing the interpretation of the community from a purely $9-11$ attack to a mix with the attack to Afghanistan. We note how this switch also happens due to the necessity of having 4 communities per timestep, and the appearance of `anthrax' implied the disappearance of one community %(we decided to keep `attack' over `political' due to its bigger weight from weeks 3 to 4, but one could similarly do the contrary). 
These same words move from week 5 to the `Afghanistan' community.

%\indent $\quad \bullet$  
Community `Afghanistan' during the first three weeks is composed by the words related to the Taliban word (\textit{Afghanistan}, \textit{Taliban}, \textit{Bin Laden} and \textit{Muslim}) and after weeks 4 and 5 it grows including the aforementioned words related to the war in Afghanistan.

The community `political' is represented by words explaining the political scene, which however is blended with the military and the financial spheres. Some representatives words are \textit{foreign minister}, \textit{official}, \textit{war}, \textit{terrorism}. In week 1 there is strong presence of financial words within political, such as \textit{economy} or \textit{financial}. As explained, `political' ceases to exist on its own by week 4 and gets merged with `attack'. Some of its words are absorbed in others, and at the same time a new community appears at this time, named `anthrax'. 

`Anthrax' appears when the first death due to the anthrax attacks was observed on October 5. Letters containing anthrax spores were mailed to media offices and political figures, leading to the illness of $22$ people and the deaths of $5$, according to the Federal Bureau of Investigation (FBI) and the Centers for Disease Control and Prevention (CDC). Top words of are \textit{case}, \textit{antibiotic}, \textit{spore}, \textit{test}, \textit{bacterium}, \textit{news}, \textit{anthrax}, \textit{letter}, \textit{deadly}. Note that also one of the top words is \textit{Brokaw}, which is the name of an American NBC journalist who covered the event and also received a letter with the spore of anthrax. In weeks $5$ and $6$ we observe some words switching across `anthrax' and `security' and vice versa (\textit{official}, \textit{health}, \textit{news}, \textit{senate} which indicate the dual nature of those communities that both relate to health security. In the first weeks, `security' is mostly related to air and state security, including words such as \textit{passenger}, \textit{airline} and \textit{federal}, \textit{agent}, \textit{FBI}. In week 3 this community evolves around the event involving the journalist Yvonne Ridley, who was held captive by the Taliban in Afghanistan after the attacks but was later released and returned to the UK.
%\textcolor{red}{The difference between 4 and 5 is that the political splits into political and war on terror? If that's the case then it's def worth it having results for both p=4 and p=5.}

The Sankey diagram we can only represent thresholded affiliations to single communities, ignoring the overlap that our model allows. In order to see the multiple memberships for each word used we refer to Figure~\ref{fig:pie_plots_reuters}. For example, \textit{president Bush} and \textit{war} switch in weeks 3, 4 and 5 from `political response' to `attack' and then `Afghanistan' as highlighted previously. From the pie plots, we can also appreciate how hard clustering is a crude representation of reality, since many of these words have a more nuanced belonging to various communities: \textit{city}, which is affiliated to various communities, switches meaning by being first associated mostly to New York and then to the city of Kandahar in Afghanistan, one of the principal site of the US bombing in October 2001. In \textit{letter}, \textit{office}, \textit{senate} and \textit{health} we see both affiliations to `anthrax' and `security'.

\subsection{Comparisons}
\subsubsection{Sparse network with overlapping communities (SNetOC)} 
As a baseline, we fit the static overlapping community model of~\cite{todeschini2020exchangeable} (SNetOC) on an aggregated static view of the graph, using the code provided by the authors. This model is designed to handle well sparsity and degree heterogeneity with overlapping community structure. %We aim to show that incorporating temporal dynamics explains the data better than this static approach. 
We run the model with $p=5$ communities overall. %As expected, the model gives a good fit on the degree distribution. 
 We get point estimates of the weights using the minimum Bayes risk point estimate as done in the original paper and assign the words to their highest membership to get the communities as shown in Table~\ref{tab:static_reuters_p5}. %There are overall $122$ words in community 1,  $166$ in 2, $152$ in 3, $256$ in 4 and $170$ in 5. 
%In general there is a meaningful recovery of communities and communities contain words of various degrees. 
Taking the top representative words of each community we see that communities `anthrax', `attack' and `political' are quite similar in meaning as in our case. We renamed our `Afghanistan' and `security' communities as `Taliban' and `war on terror' due to the words belonging to them. 
\begin{table}[t!]
\caption{Top representative words for each community under SNetOC with aggregated data from 6 weeks with $p=5$.}
\centering
\begin{tabular}{|c|c|}
\hline
{Community} & {Representative words} \\ 
\hline
`attack’ &  plane, hijack, WTC, flight, airport, passenger,  \\ & September,
 hijacker, jet, tower, pentagon, pilot
\\ 
\hline
`war on terror’ & US, attack, Afghanistan, Bush, people, country, terrorism, war, Bin Laden, \\ &
military, Washington, force, Al Quaeda, state, campaign, muslim, action \\ 
\hline
`political’  & security, bank, foreign, minister, airline, health, \\ & financial, business, Australian, committee, trade, German  \\ 
\hline
`anthrax’  & anthrax, official, letter, New York, mail, test, office, building, news, \\ & case, capitol, FBI, bacterium, senate, antibiotic,
 post, hospital 
\\ 
\hline
`Taliban’ & Taliban, Kabul, city, Kandahar, tell, capital, reporter, Pakistani, mullah, \\ & southern, opposition, border, province, resident, Israeli
\\
\hline
\end{tabular}

\label{tab:static_reuters_p5}
\end{table}
Using this static approach we cannot see words change communities, how the theme of a community evolves, appears or disappears, making it not possible to track any dynamic evolution. For example in our model, `attack' is the `9-11' attack but later on it acquires additional words regarding the attack in Afghanistan, whereas in the static model `attack' only concerns words of `9-11 attack'. Another example is the community `security' which in our model first is about airport and airline security whereas later relates to health and state security. In the static case, we cannot see this evolution at all and in fact, security does not exist as a community on its own as it is blended with the rest. 
\subsubsection{Sparse network with overlapping communities (SNetOC) at each week}
\begin{figure}[t!]
\centering
\includegraphics[width=12cm]{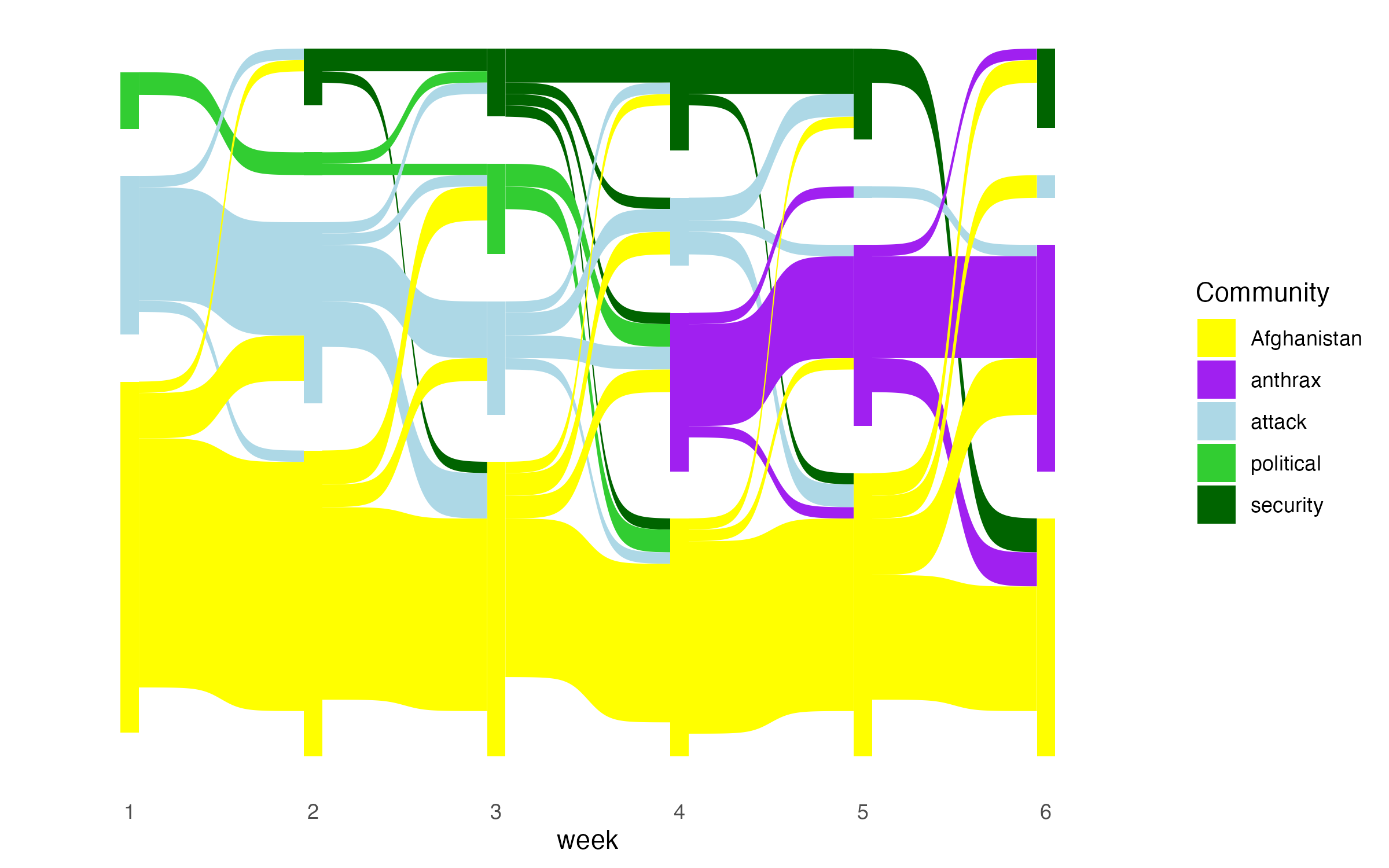}
\caption{Sankey plot of weekly affiliation of representative words using SNetOC at each timestep with $p=4$.}
\label{fig:sankey_plot_reuters_tode}
\end{figure}
{We run SNetOC at each timestep with $p=4$. %We interpret the communities on the basis of the minimum Bayes risk point estimate as done in the original paper. 
As expected, the posterior predictive degree distribution is good. There is a meaningful recovery of communities, but the model cannot track their evolution, since we have to estimate $T$ different models. For example the words \textit{war, Bush, terrorism, US} are throughout in community `Afghanistan' and do not belong to `political' as they do under our model (for $t=1,2,3$). This shows that there is no meaningful mixing between these two communities. Similarly, there is less mixing between `security' and `anthrax' as the words \textit{anthrax, health, letter} are affiliated with `anthrax' but they do not pick up a `security' component as they do in our case.} 
The pie plots of the words are found in %Figure~\ref{fig:pie_plots_reuters_tode} in 
Section B3 %~\ref{sec:B3}
of the Supplementary. The sankey plot is shown in Figure \ref{fig:sankey_plot_reuters_tode}.
It is evident that $p=4$ is not the best choice. Indeed at week 1 there are only 3 communities, one of which (`political') contains only 5 words, of which only 2 (\textit{foreign} and \textit{minister}) survive over the subsequent weeks. The vast majority of words belongs to the `attack' community, which in the first 2-3 weeks is about the WTC attack. 
\begin{figure}[t!]
    \centering
    %\subfigure[t=0]
    {\includegraphics[width=5cm]%{figures/reuters/degree_distr_t0_non_hier_dSBM.png}}
    {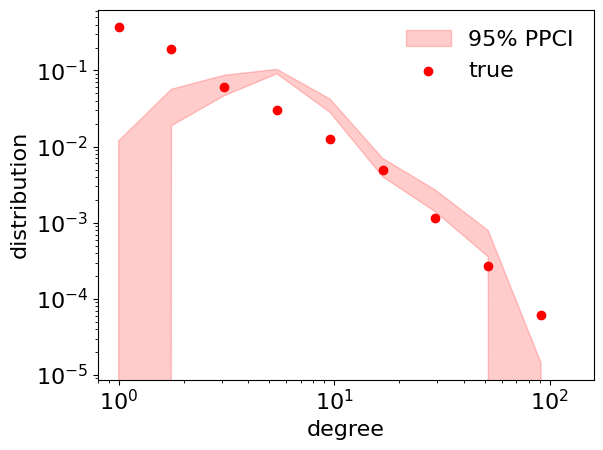}}
    {\includegraphics[width=5cm]%{figures/reuters/degree_distr_time_1_non_hier_dSBM.png}}
    {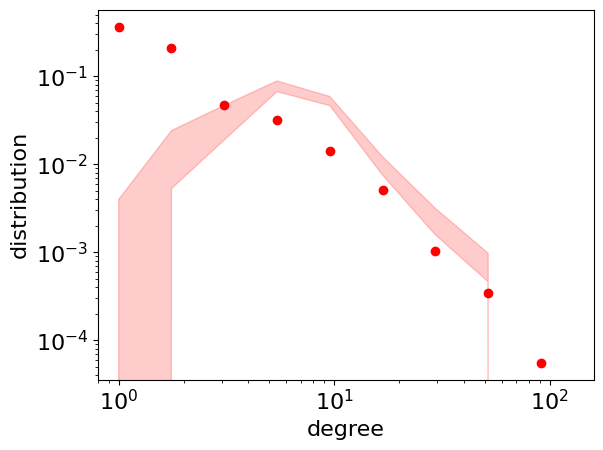}}
    {\includegraphics[width=5cm]%{figures/reuters/degree_distr_time_2_hier_v2.png}}
    {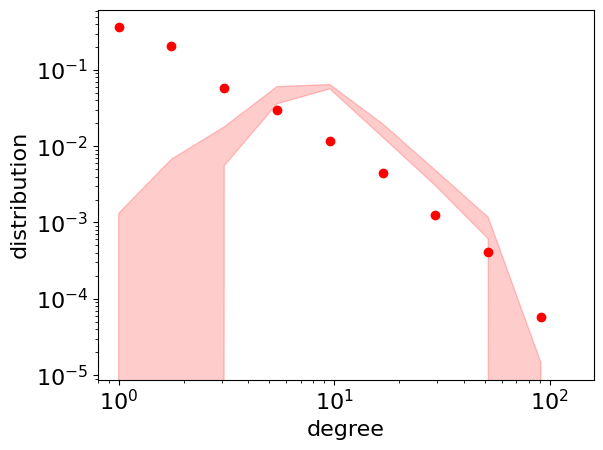}}
    {\includegraphics[width=5cm]%{figures/reuters/degree_distr_time_3_non_hier_dSBM.png}}
        {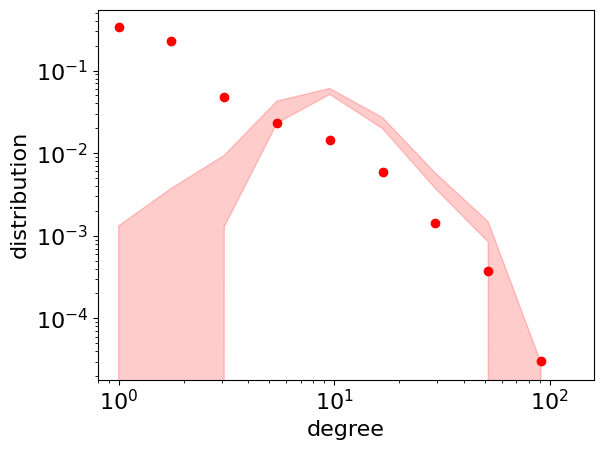}}
    {\includegraphics[width=5cm]%{figures/reuters/degree_distr_time_4_non_hier_dSBM.png}}
    {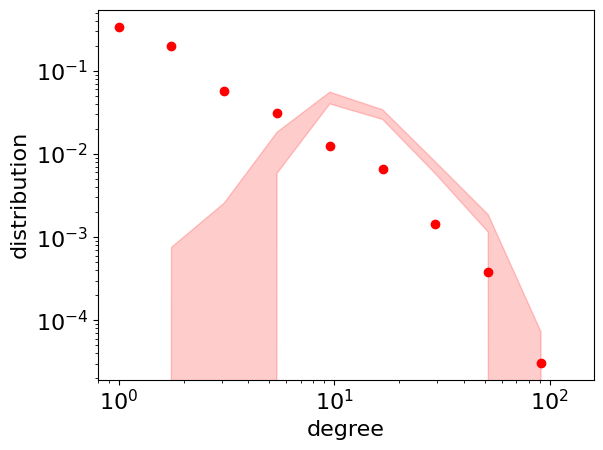}}
    {\includegraphics[width=5cm]%{figures/reuters/degree_distr_time_5_non_hier_dSBM.png}}
    {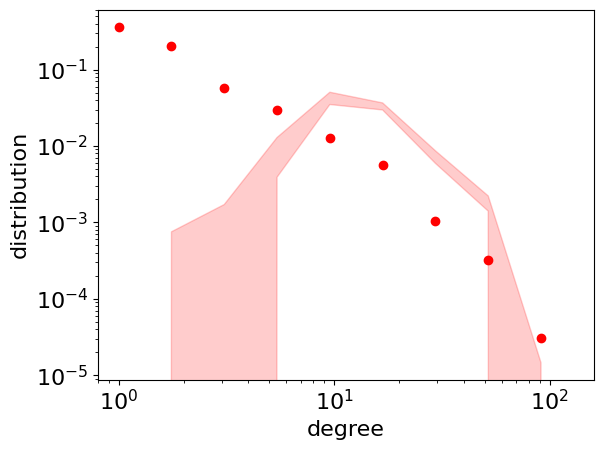}}

    \caption{Degree distribution for dynSBM in log-log scale, empirical in red dots and 95\% PPCI in shaded region for $t=1,2,3$ (top row), and $t=4,5,6$ (bottom row).}
\label{fig:dynSBM_degree_distr_v2}
\end{figure}
The rest of the words are in `Afghanistan', which has a broader meaning than it had with our dynamic proposal, including words from the sphere of Taliban, terrorism, war, US (\textit{United States, Bush, national}) and the political/financial worlds (\textit{security, financial, coalition}). As time evolves, `Afghanistan' and `attack' tend to shrink and new communities arise, such as `security' and `anthrax'. `Security' is a combination of national security coming from the attack to Afghanistan (\textit{Taliban, foreign, Kabul, Kandahar, national, Pentagon}), airport security (\textit{airport, plane}) and health security (\textit{hospital}). In the last two weeks, again the results seem to suggest that $p=3$ or $p=2$ would provide a better fit. 

Overall, SNetOC on the aggregated data from all timesteps provides interesting results for $p=5$,  but these do not carry out to similarly interpretable results when we separate the timesteps. 
It is evident that our proposal, which takes into account time dynamics, is able to better represent the evolution of communities through different meanings across timesteps. For example, through a semantic change of `attack' from WTC to the attack to Afghanistan, or the change of `security' from airport security to national security and then health security. With SNetOC, the `attack' community loses strength over time, and `security' appears in a weaker form (less words and without a clear semantic change, on the contrary a mix of concepts from the beginning).
\begin{figure}[t!]
    \centering
    %\subfigure[t=0]
    {\includegraphics[width=5cm]{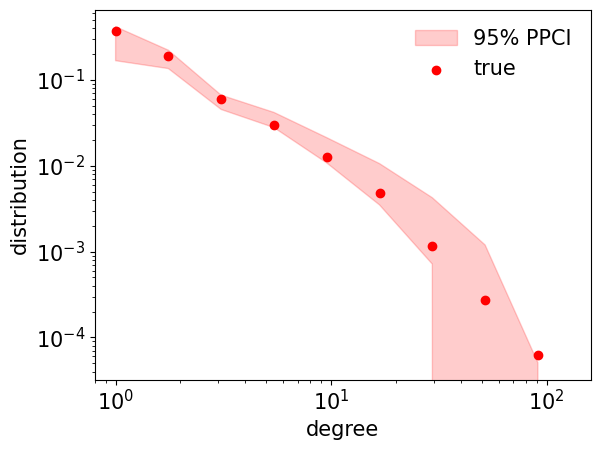}}
    %\subfigure[t=1]
    {\includegraphics[width=5cm]{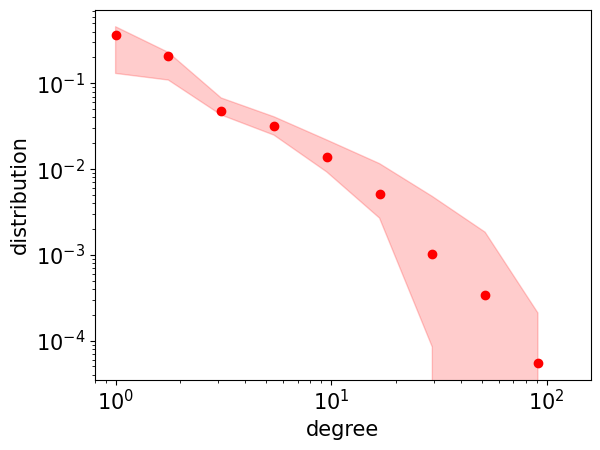}}
    %\subfigure[t=2]
    {\includegraphics[width=5cm]{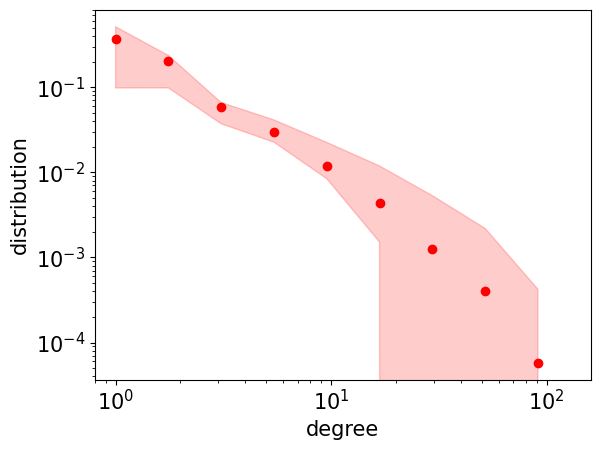}}
    %\subfigure[t=3]
    {\includegraphics[width=5cm]{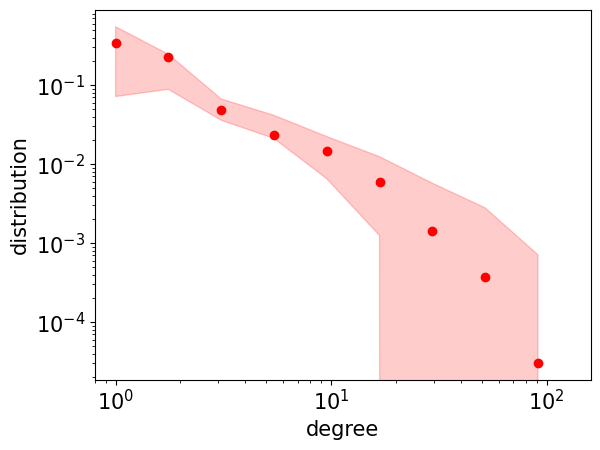}}
    %\subfigure[t=4]
    {\includegraphics[width=5cm]{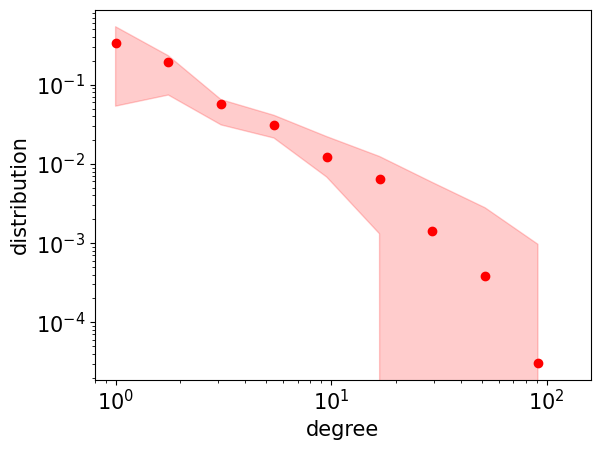}}
    %\subfigure[t=5]
    {\includegraphics[width=5cm]{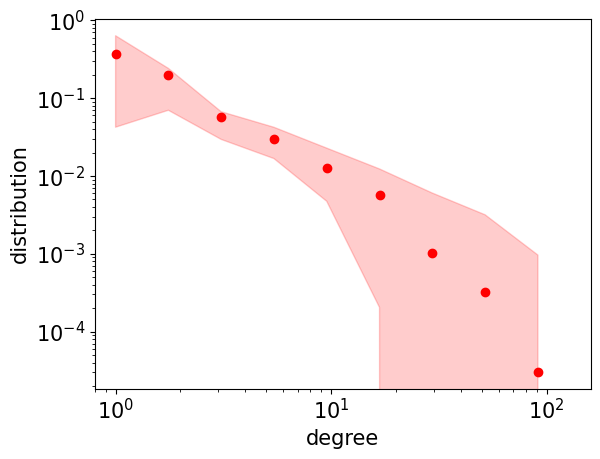}}
    \caption{Degree distribution in log-log scale for dMMSB, empirical in red dots and 95\% PPCI in shaded region for $t=1,2,3$ (top row), and $t=4,5,6$ (bottom row).}
\label{fig:dMMSBM_degree_distr_PPCI}
\end{figure}
\subsubsection{Dynamic Stochastic Blockmodel}
The Dynamic Stochastic Blockmodel (dynSBM)~\citep{matias2017statistical} is the celebrated temporal extension of the SBM where nodes have dynamic single memberships. The formulation of the model is found in Section B4 %~\ref{sec:B4}
of the Supplementary. Parameter estimation is performed with a 
variational Expectation–Maximization (VEM) algorithm implemented in R~\citep{matias2017software}.
We run dynSBM on the binary graph with $p=4$ % using the software  %We use the binary graph as their approach does not allow multiedges in the way we do. 
and %use the point estimates of the parameters %and the (hard) cluster memberships of the nodes (a single community per node) 
generate graphs from the posterior predictive. %What we can do is to see how the nodes are clustered. We expect that they are clustered based on their connectivity?
As shown in Figure~\ref{fig:dynSBM_degree_distr_v2}, the fit is bad, the model cannot capture well the degree distribution, especially for the low and high degree nodes. %\footnote{We cannot provide similar uncertainty intervals for dynSBM in the same way as in our case.} %intermediate values in the VEM are optimization steps and not sampling steps.} The model does not fit well the sparse, power- law setup and the degree distribution from graphs arising from the estimated parameters is far from the true one as Figure~\ref{fig:dynSBM_degree distr} shows. The model fails to model the low degree nodes such as the degree 1 ones which is important part of our purposes (nodes with degree 1 are related to sparsity).
Regarding the discovered communities, there is no meaningful thematic interpretation. Interestingly, at all timesteps, one of the communities has around $80\%$ of the nodes. We believe that the poor fit is caused by the high sparsity of the connections and the scale-free degree distribution, which are not features handled by the dynSBM.
%and these nodes are of average and low degree - specifically none of the high degree nodes ends up here. The high degree nodes are always split into two specific communities. These two communities also contain some other medium and lower degree nodes and together form around $15\%$ of the nodes. The final community contains the remaining $5\%$ of the words which are only low degree nodes. For this and the next mode, we ran the model $10$ times with different random seeds and verified that the results remained consistent.}

\subsubsection{Dynamic Mixed Membership Stochastic Blockmodel}
The Dynamic Mixed Membership Stochastic Blockmodel (dMMSB)~\cite{xing2010state} is the extension of the SBM where nodes have dynamic mixed group memberships.
The formulation is in Section B5 %~\ref{sec:B5}
of Supplementary. 
% \textcolor{red}{What is the mathematical formulation of the model}? The model is similar to dynSBM but here, each node's latent group affiliation is described as $Z_i^{(t)} \in [0, 1]^p$, this formulation allows $Z$ to act as membership matrix, modeling the probability of a node $i$ belonging to each community. These affiliations evolve over time based on a state-space model instead of a markov chain as in dynSBM. \textcolor{red}{TODO: add final likelihood + cleanup}. 
Given the lack of software, we created code\footnote{\url{https://anonymous.4open.science/r/dMMSB-09B1/README.md}} to implement the proposed EM algorithm. 
We run dMMSB on the binary graph with $p=4$ and generate graphs from the posterior predictive, shown in Figure~\ref{fig:dMMSBM_degree_distr_PPCI}. The fit is better than dynSBM, but this comes at expense of the community structure. In fact, the block matrix is estimated to have a single non-zero entry on the diagonal with all off-diagonal elements set to zero, implying that all connections happen inside a single community. Hard-clustering the nodes shows that the only information used to assign words to communities is the degree: the only active community is made of the highest degree nodes. 
%Instead, if any variation exists in $\pi$, it acts merely as a proxy for sociability, rather than latent community structure. 
Overall, the estimated model fails to function as a community detection algorithm for this sparse dataset and only manages to model the degree heterogeneity of the data by using the communiy parameters. 
%Regarding community interpretation, nodes are clustered according to their degree. At all timesteps there is a similar pattern of adding high degree words in one community, and then the low degree ones are split into those with around 1-5 connections and then 5-15 to another community, leaving the last community almost empty (2-3 nodes or less). This shows the weakness of the model to recover well both the degree heterogeneity and meaningful communities. %\textcolor{red}{Alternative Explanation:} 
 %\textcolor{red}{We ran the model 10 times with different random seeds and verified that the results remained consistent.}

%
%
\section{Conclusion}
\label{sec:conc}
We propose a Bayesian nonparametric model for dynamic evolution of communities in sparse graphs with power-law degree distribution. In this way, our work generalizes existing overlapping community models to the sparse and power-law regime and its applicability is demonstrated on a real-world graph. Future work could be the inclusion of covariates or an estimation of the number of communities. 

\textbf{Acknowledgements and Contributions} AL implemented and applied the code of dMMSB, contributed to our model's code and performed the reuters' data cleaning. Everything else was done equally by XM and FP. The authors would like to thank Professors François Caron and Chryssis Georgiou for the insightful discussions. 
XM has received funding from the European Union's Horizon programme under Marie Skłodowska-Curie for the project “DyNeMo”, id 101151781 and from the University of Cyprus, startup grant programme.
FP has received funding from the Sapienza University of Rome for the project “Bayesian models for sparse networks and applications” grant number RP1241910EF9AF1F.
  
\bibliography{JASA_paper/bnpnetwork_MM-MC}
%need to fix biblio

\newpage
\spacingset{1.8} % DON'T change the spacing!
\begin{center}
    \Large Supplementary for Dynamic Sparse graphs with overlapping communities
\end{center}

%\section*{A}
\subsection*{A1. Proof of propositions 2 and 3}
\label{sec:A}
%\textbf{Proof of Propositions 2 and 3}
\begin{proof}
Propositions %2 %~\eqref{prop:sparsity} 
%and 3 %\eqref{prop:deg_distr} 
follow from Remark 5 in \cite{caron2023sparsity} upon noting that 
\begin{align}
    1-e^{-2 \sum_{k=1}^p w_{ki}^{(t)} w_{kj}^{(t)}} = 1-e^{-2w_{0i}w_{0j}\sum_{k=1}^p\beta_{ki}^{(t)} \beta_{kj}^{(t)}}= 1 -e^{\eta(w_{0i},w_{0j})\omega(\beta_{i}^{(t)},\beta_{j}^{(t)})}
\nonumber
\end{align}
where $\omega(\beta_{i}^{(t)}, \beta_{j}^{(t)}):=\sum_{k=1}^p\beta_{ki}^{(t)}\beta_{kj}^{(t)}$ and $\eta(w_{0i},w_{0j}):=2w_{0i}w_{0j}$. To map back to the notation of \cite{caron2023sparsity} we note that $w_{0i}$ can be obtained as $w_{0i}=\Bar{\rho_0}^{-1}(\vartheta_i)$, with $\rho_0$ the L\'evy measure of the generalized gamma process, $\Bar{\rho_0}(x)=\int_x^\infty \rho_0(dy)$ its tail L\'evy intensity and $(\vartheta_i)_{i\ge 1}$ random variables sampled from a unit-rate Poisson process. 
\end{proof}

\subsection*{A2. Posterior distribution}
\label{sec:A2}
%\textbf{Posterior Distribution}

The full posterior at $t$ is
%\label{eq:approx_posterior_detailed}
\begin{align}
    \log p(\mathbf w_0, &\boldsymbol{\beta}^{(t)}_{1}, \dots, \boldsymbol{\beta}^{(t)}_{p}, \boldsymbol{\gamma}^{(t)}_{1}, \dots, \boldsymbol{\gamma}^{(t)}_{p},
    \boldsymbol{\gamma}^{(t-1)}_{1}, \dots, \boldsymbol{\gamma}^{(t-1)}_{p}, \xi \mid \mathbf n^{(t)}) \notag\\
    \propto & \log p(\mathbf n^{(t)}|\mathbf w_0, \boldsymbol{\beta}^{(t)}_{1}, \dots, \boldsymbol{\beta}^{(t)}_{p}) + \log p(\mathbf w_0)
    %+ \log p(\mathbf u|\mathbf w_0) 
    + \sum_{k}\log p(\boldsymbol{\beta}^{(t)}_{k}|\boldsymbol{\gamma}^{(t)}_{k}, \boldsymbol{\gamma}^{(t-1)}_{k}) + \log p(\xi) \notag\\ 
    \propto  & \left[\sum_{i=1}^L \sum_{j=1}^L \log\text{ Poisson }\left({n^{(t)}_{ij};w_{0i}w_{j0}\sum_k \beta^{(t)}_{ki} \beta^{(t)}_{jk}}\right)\right]\notag\\
        & + \left[\sum_{i=1}^L \left(\log\text{BFRY}(w_{0i}; \alpha/L,\tau,\sigma) 
    %+ \log\text{tExp}(u_{i};w_{0i},t_{\alpha,\sigma}) 
    + \sum_k\log p\left(\beta^{(t)}_{ki}|\gamma^{(t-1)}_{ki}, \gamma^{(t)}_{ki}\right)\right) \right]+ \log p(\xi)\notag\\ 
    %& + \log\text{Gamma}(\sigma; a_\sigma, b_\sigma) + \log\text{Gamma}(\tau; a_\tau, b_\tau) + \log\text{Gamma}(\alpha; a_\alpha, b_\alpha) + \log\text{Gamma}(\psi; a_\psi, b_\psi)\notag\\
    %next line
    \propto  & \sum_{i=1}^L \Biggl( m_{ti}\log \left(w_{0i}\sum_k\beta_{tki}\right) - (1+\sigma)\log w_{0i} - \tau w_{0i}
    + \log(1-e^{-(\sigma/\alpha)^{1/\sigma}w_{0i}})
    \notag\\
    &+\sum_{k=1}^p \left(a_k+\psi-1+\mathbbm{1}_{t>1}\psi\right) \log \beta^{(t)}_{ki}- \left(b_k+\gamma^{(t)}+\mathbbm{1}_{t>1}\gamma^{(t-1)}_{k}\right)\beta^{(t)}_{ki} \notag\\
    &+(a_k+\psi+\mathbbm{1}_{t>1}\psi) \log(b_k+\gamma^{(t)} +\mathbbm{1}_{t>1}\gamma^{(t-1)}_{k}) -\Gamma(a_k +\psi+\mathbbm{1}_{t>1}\psi )  \Biggr)\notag\\
    & + \log\sigma -\log \Gamma(1-\sigma) - \log\left(\left(\tau + t_{\alpha,\sigma}\right)^{\sigma} -\tau^{\sigma}\right)\ - \left(\sum_{i=1}^L w_{0i}\sum_k\beta_{tki}\right)^2 \notag \\
    & + \log p(\sigma) + \log p(\alpha) + \log p(\tau) + \log p(\psi)+ \log p(a)+\log p(b), 
%%+ (a_\sigma-1)\log \sigma-b_\sigma \sigma + (a_\tau-1)\log\tau-b_\tau \tau \\
%%&\left.\qquad + (a_\alpha-1)\log\alpha-b_\alpha\alpha +(a_\psi-1)\log\psi-b_\psi\alpha
\end{align}

where in the last line we have the (independent) priors of the hyperparameters.
\newpage
%\subsection*{B}
\subsection*{B1. MCMC trace plots for synthetic data}
\label{sec:B1}
%\textbf{}
\begin{figure}[H]
    \centering
    %\subfigure[$\alpha$]{\includegraphics[width=5cm]{figures/synthetic/MCMC_trace_alpha.png}}
    %\subfigure[$\log (\alpha \tau^\sigma)$]
    {\includegraphics[width=5cm]%{figures/synthetic/MCMC_plot_logalphatausigma.png}}
{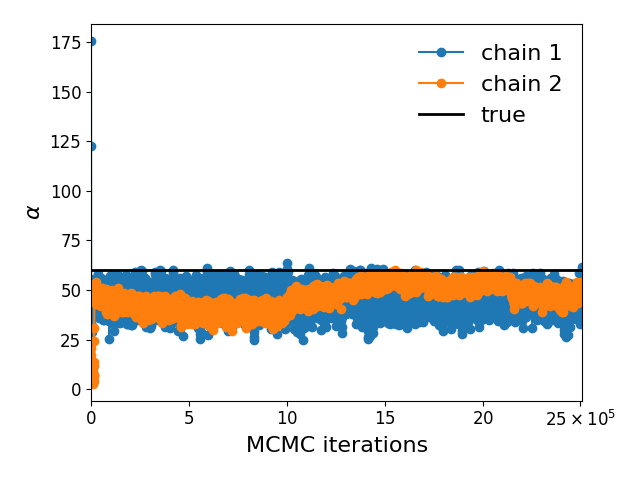}}    
    %\subfigure[$\sigma$]
    {\includegraphics[width=5cm]{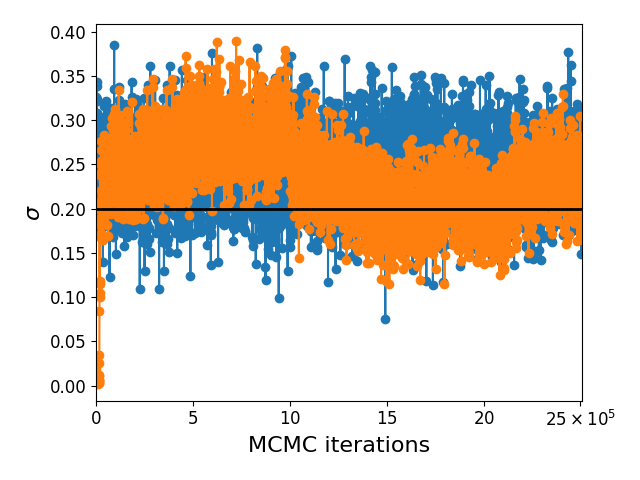}}
    %\subfigure[$\psi$]
    {\includegraphics[width=5cm]{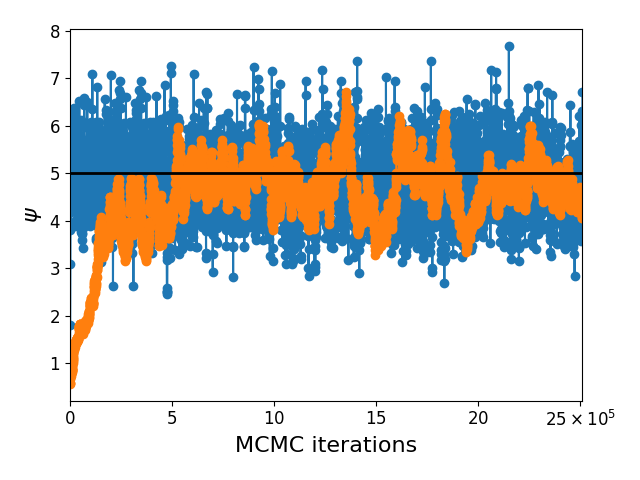}}
    %\subfigure[$a_1$]
    {\includegraphics[width=5cm]{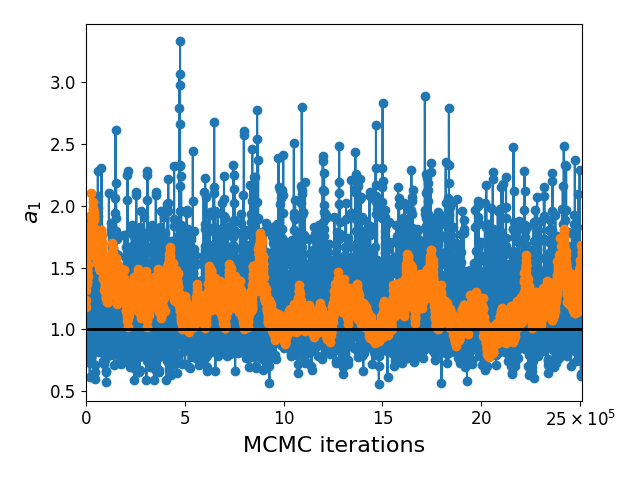}}
    %\subfigure[$a_2$]
    {\includegraphics[width=5cm]{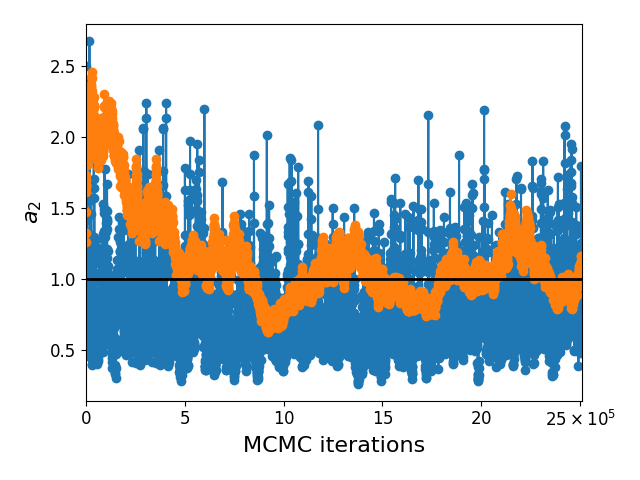}}
    %\subfigure[$b_1$]
    {\includegraphics[width=5cm]{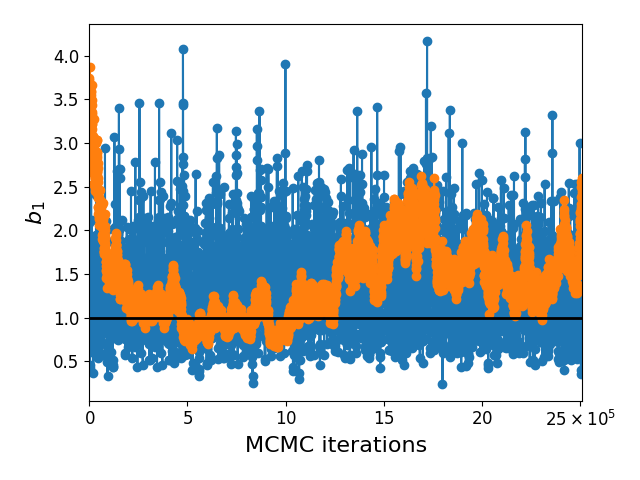}}    %\subfigure[$b_2$]
    {\includegraphics[width=5cm]{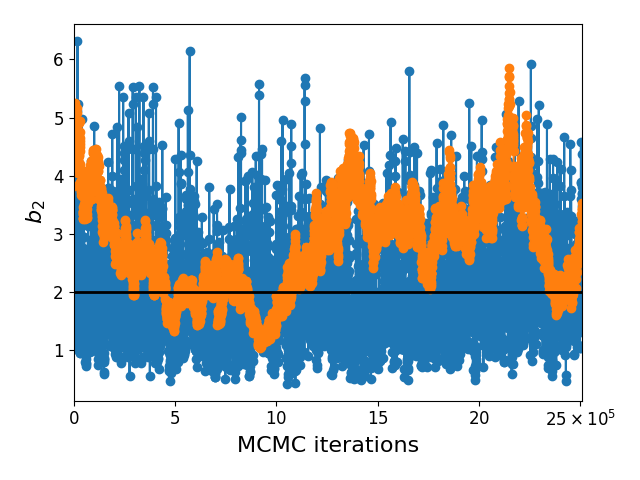}}
    %{\includegraphics[width=5cm]{figures/synthetic/MCMC_logposterior.png}}
    \caption{MCMC trace plots for the parameters on the synthetic dataset.}
\label{fig:trace_plots_synthetic}

\end{figure}

\begin{comment}
\begin{figure}[H]
    \centering
{\includegraphics[width=.5\textwidth]{figures/synthetic/MCMC_logposterior.png}}
    \caption{Log-posterior of the true model (black) and from the chain 1 and 2 targeting the approximate posterior (blue, orange) \textcolor{red}{find true posterior and fix legend}}
\label{fig:log_posterior_synthetic}
\end{figure}
\end{comment}

\newpage
\subsection*{B2. MCMC trace plots for Reuters news data}
\label{sec:B2}
%\textbf{MCMC trace plots for Reuters News data}
\begin{figure}[ht]
    \centering
    %\subfigure[$\log \alpha \tau^\sigma$]
{\includegraphics[width=5cm]{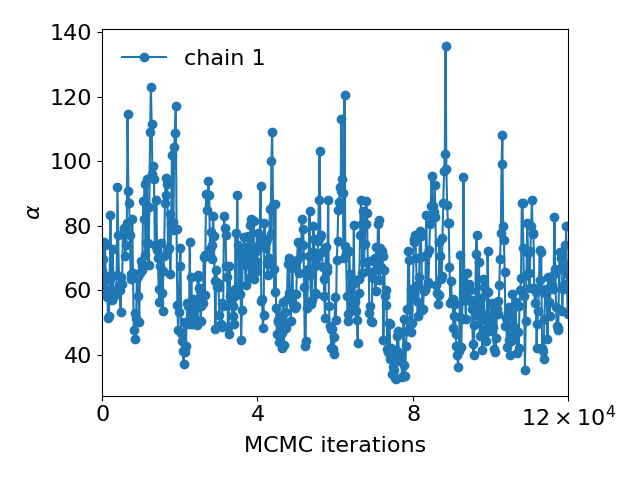}}
%\subfigure[$\sigma$]
{\includegraphics[width=5cm]{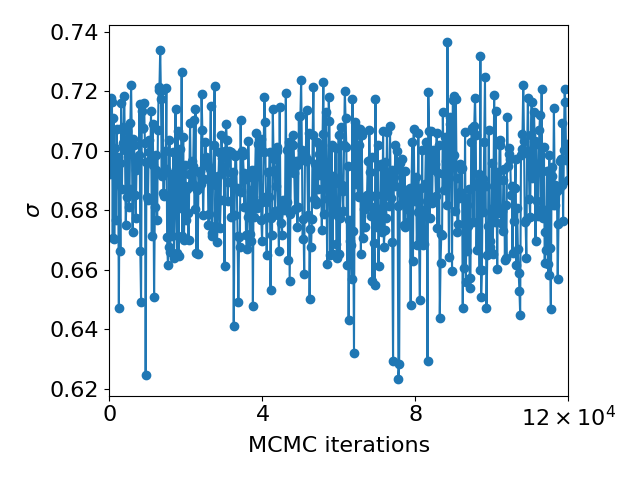}}
    %\subfigure[$\psi$]
{\includegraphics[width=5cm]{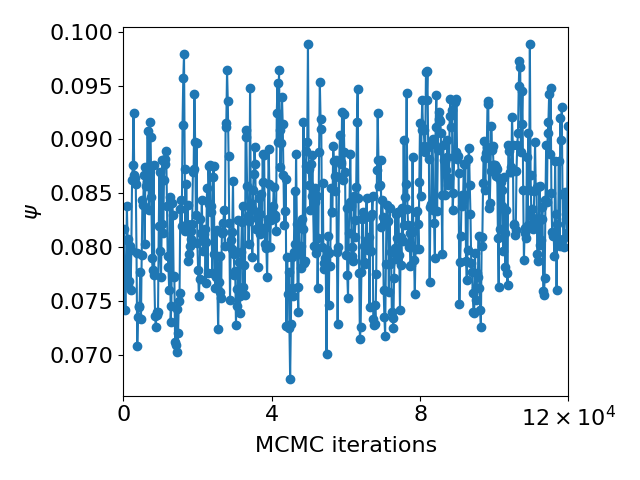}}
    %\subfigure[$a_1$]
{\includegraphics[width=5cm]{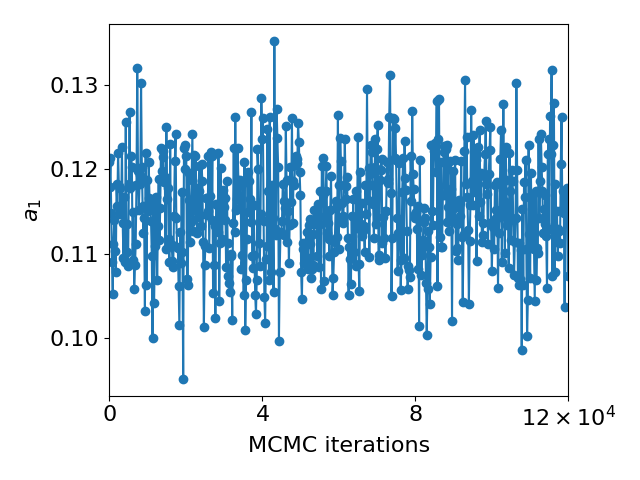}}
    %\subfigure[$a_2$]
{\includegraphics[width=5cm]{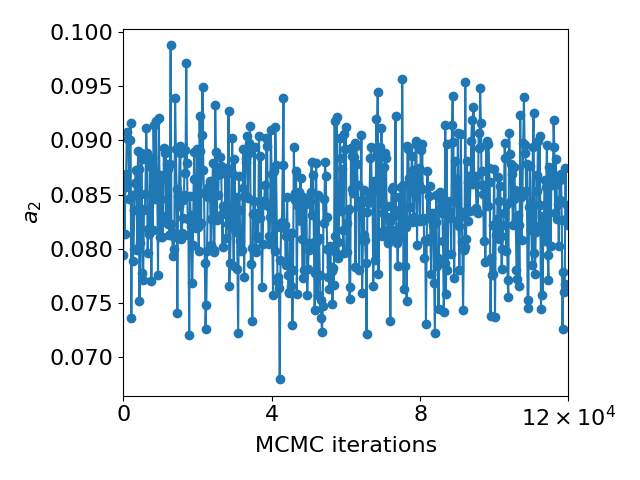}}
     %\subfigure[$a_2$]
{\includegraphics[width=5cm]{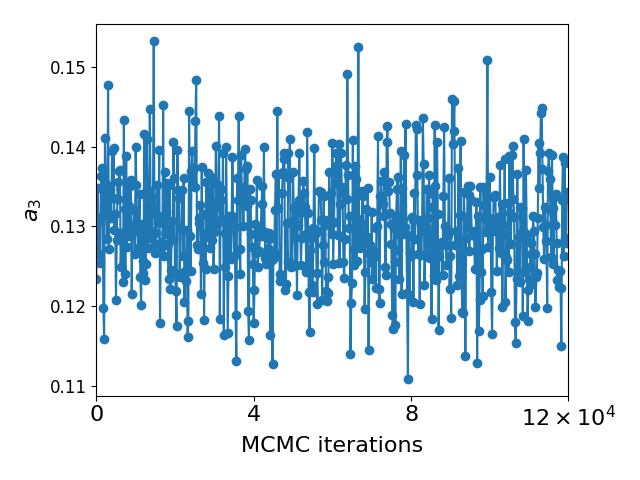}}
    %\subfigure[$a_2$]
{\includegraphics[width=5cm]{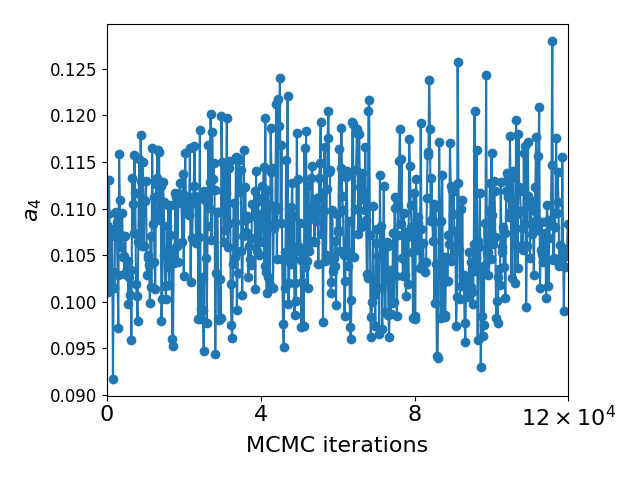}}
    %\subfigure[$a_2$]
%\subfigure[$b_1$]
{\includegraphics[width=5cm]{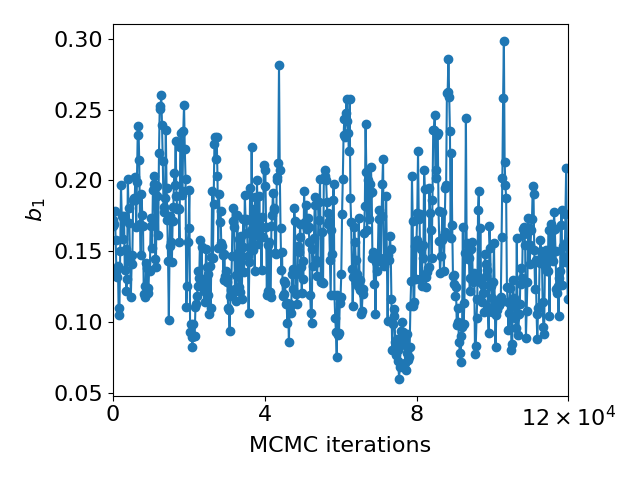}}
% \subfigure[$b_2$]
{\includegraphics[width=5cm]{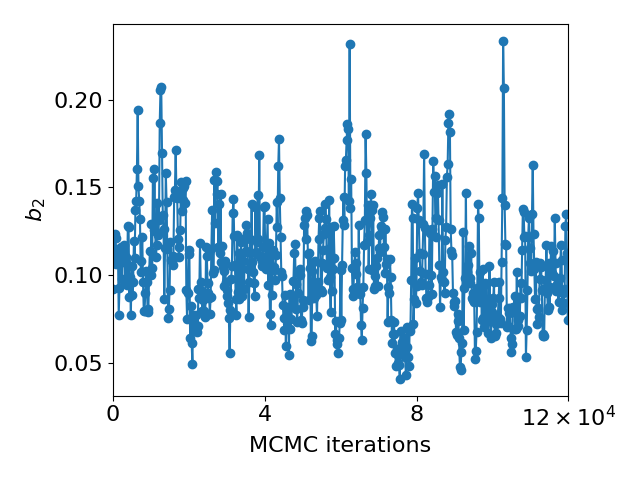}}
% \subfigure[$b_2$]
{\includegraphics[width=5cm]{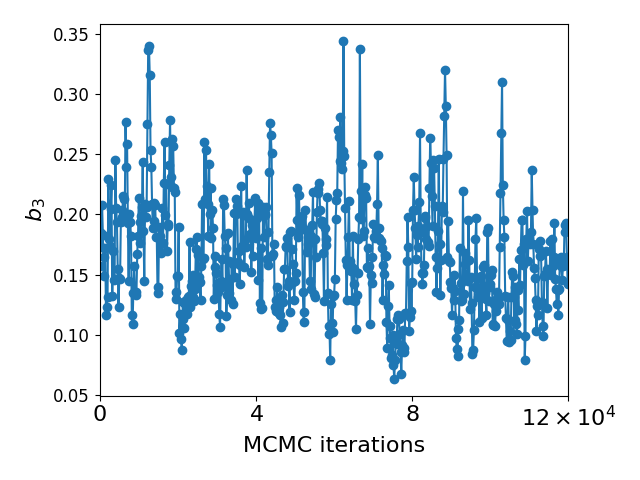}}
% \subfigure[$b_2$]
{\includegraphics[width=5cm]{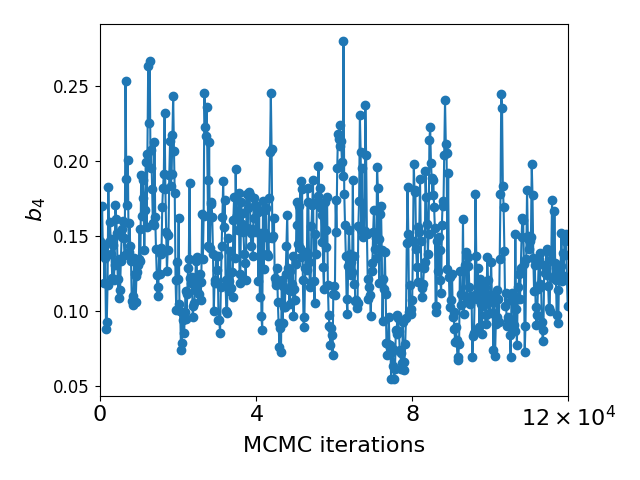}}
%
%{\includegraphics[width=5cm]{figures/reuters/logPosterior2.png}}
\caption{MCMC trace plots of the parameters for the reuters dataset.}    
\label{fig:trace_plots_reuters}
\end{figure}

\newpage 

\newpage
\subsection*{B3. Pie charts of weekly affiliations from SNetOC per time step}
\label{sec:B3}

\begin{figure}[H]
    \centering
    %\begin{minipage}{.45\textwidth}
    \includegraphics[height=16cm]{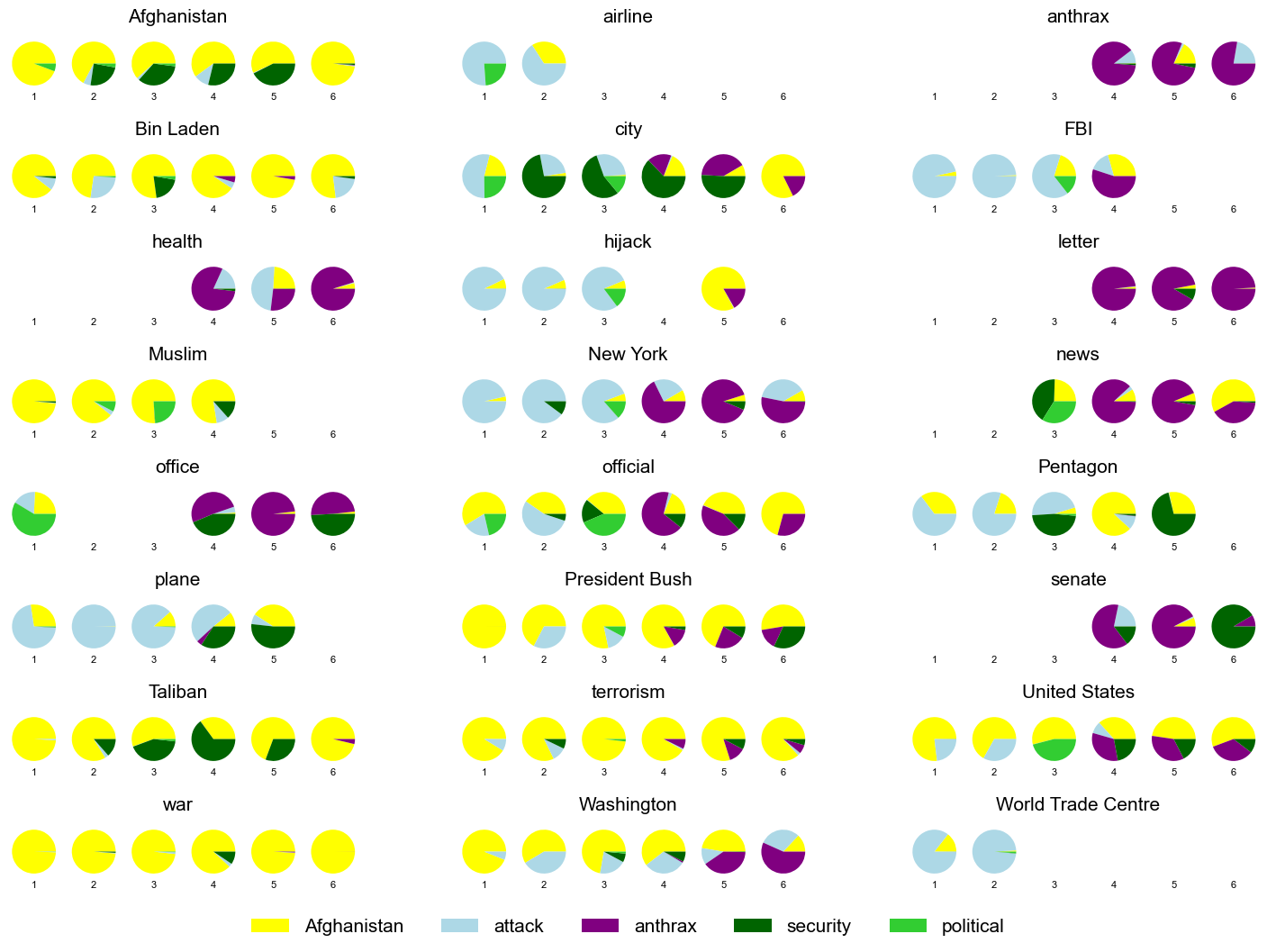}
    %\end{minipage}
    %\begin{minipage}{.45\textwidth}
    %\includegraphics[width=7cm]{figures/scores_p5_pie_plot_listwords.png}
    %\end{minipage}
    %\caption{Pie plots of weekly affiliation of principal words. On the left, for $p=4$, on the right $p=5$.}
    \caption{Pie charts with the \% of community affiliations for some high degree words at each timestep, using SNetOC at each timestep with $p=4$.}
\label{fig:pie_plots_reuters_tode}
\end{figure}

\subsection*{B4. Model formulation for dynSBM}
\label{sec:B4}

Let $\{G_t = (V, E_t)\}_{t=1}^T$ denote a sequence of graphs on a set of $N$ nodes, 
with symmetric adjacency matrices $Y^{(t)} = [Y^{(t)}_{ij}]_{i,j=1}^N$ (with no self-loops). 
Each node $i$ is associated with a latent group label $Z^{(t)}_{i} \in \{1, \dots, p\}$ at time $t$, 
which evolves according to a first–order Markov chain:
$ \Pr(Z^{(t)}_{i} = q \mid Z^{(t-1)}_{i} = r) = \Xi_{rq}, 
$
with $ \Xi $ the $p \times p$ transition matrix and $\zeta=(\zeta_1,\dots,\zeta_p)$ its initial stationary distribution.
%$ \Pi \in [0,1]^{Q \times Q}, \quad \sum_{q=1}^Q \Pi_{rq} = 1.$
Conditional on the latent memberships at time $t$, edges are independent with
$
Y^{(t)}_{ij} \mid (Z^{(t)}_{i}=q,\, Z^{(t)}_{j}=\ell) \sim \mathrm{Bernoulli}\big(\beta^{(t)}_{q\ell}\big),$
where $\beta^{(t)}_{q\ell} \in [0,1]$ denote the connection probabilities 
between groups $q$ and $\ell$ at time $t$.Note that the authors also propose ways to treat weighted graphs but none of their approaches includes our Poisson multiedges, hence we need to resort to the binary symmetric graphs.

\subsection*{Β5. Model formulation for dMMSB}
\label{sec:B5}

Let $\mu^{(t)} = (\mu_1^{(t)}, \dots,\mu_p^{(t)})$ be a mixed membership prior that follows a state-space model $\mathrm{Normal}(\mathbf{A}\mu^{(t-1)}, \Phi)$ initialized at $t=1$ from $\mathrm{Normal}(\nu, \Phi)$. %The interaction probabilities between roles evolve in parallel. 
For each pair of roles $r, q$ an underlying $\eta_{rq}^{(t)}$ compatibility parameter is introduced. At t = 1 these are drawn from $\mathrm{Normal}(\iota, \psi)$, and for $t \ge 2$ they follow a state-space model $\eta_{rq}^{(t)} \sim \mathrm{Normal}(\beta\eta_{rq}^{(t-1)}, \psi)$. These are mapped to probabilities via the logistic transform $\beta_{rq}^{(t)}=\exp(\eta_{rq}^{(t)})/(\exp(\eta_{rq}^{(t)})+ 1)$, yielding a time-indexed role-compatibility matrix $\beta_{rq}^{(t)} \in (0,1)^{p \times p}$. Edges are generated independently given the latent memberships and compatibilities. For each  $i,t$, the latent group label $Z_i^{(t)} \sim \mathrm{Multinomial}(\pi_i^{(t)}, 1)$, where $\pi_i^{(t)}$ is the mixed membership vector sampled from $\mathrm{LogisticNormal(\mu^{(t)}, \Sigma^{(t)}})$. Conditional on the latent memberships at time $t$, edges are independent with $
Y^{(t)}_{ij} \mid (Z^{(t)}_{i}=q,\, Z^{(t)}_{j}=\ell) \sim \mathrm{Bernoulli}\big(\beta^{(t)}_{q\ell}\big)$. 
This construction yields a dynamic mixed-membership network model in which both node-level memberships and role-interaction tendencies evolve over time. The variational EM algorithm described in the paper assumes that the role-compatibility matrix is constant through time and that the transition matrix $\mathbf{A}$ is equal to the identity matrix $\mathbf{I}$, this reduces the state-space model to a random walk. 

\end{document}

%% file: JASA_paper/n_2.tex
% Pie chart node with connection anchor
\begin{comment}
\newcommand{\pienodewithanchor}[6]{% name, x, y, radius, green_angle, label
  %\node[inner sep=0pt] (#1) at (#2,#3) {};
  \node[circle, inner sep=0pt] (#1) at (#2,#3) {};
  %\node[circle, draw=black, thick, inner sep=0pt, minimum size=\diam cm] (#1) at (#2,#3) {#6};
  \begin{scope}
    \fill[green!60!white] (#2,#3) -- ++(#4,0) arc[start angle=0, delta angle=#5, radius=#4] -- cycle;
    \fill[orange!80!white] (#2,#3) -- ++({cos(#5)*#4},{sin(#5)*#4}) arc[start angle=#5, delta angle={360-#5}, radius=#4] -- cycle;
    \draw[thick] (#2,#3) circle (#4);
    \node at (#2,#3) {#6};
  \end{scope}
}
\end{comment}
%% I changed the \pienodewithanchor definition to add "circle", then \draw (n1) -- (n2t) command works as it should.

% Pie chart node that is an actual circular TikZ node
\newcommand{\pienodewithanchor}[6]{% name, x, y, radius, green_angle, label
  \pgfmathsetmacro{\diam}{2*(#4)}
  % Real circular node so edges meet the border automatically
  \node[circle, draw=black, thick, inner sep=0pt, minimum size=\diam cm] (#1) at (#2,#3) {#6};
  % Draw the pie slices centered on the node
  \begin{scope}
    \fill[green!60!white] (#2,#3) -- ++(#4,0)
      arc[start angle=0, delta angle=#5, radius=#4] -- cycle;
    \fill[orange!80!white] (#2,#3) -- ++({cos(#5)*#4},{sin(#5)*#4})
      arc[start angle=#5, delta angle={360-#5}, radius=#4] -- cycle;
    % (optional) redraw the circle outline so it sits above the fills
    \draw[thick] (#2,#3) circle (#4);
    \node at (#2,#3) {#6};
  \end{scope}
}

%\begin{document}
\begin{tikzpicture}[
  roundnode/.style={circle, draw=green!60!black, fill=green!30, thick, minimum size=8mm},
  every node/.style={font=\small},
  ->, thick
]

% Arrow style that keeps arrows off the numbers (node radius ≈ 0.4cm)
%\tikzset{edge/.style={->, shorten >=0.35cm, shorten <=0.0cm}}

% --- t = 1 ---              
\pienodewithanchor{n1t1}{0}{2}{0.4}{360}{1}  % Node 1 (100% green)
\pienodewithanchor{n2t1}{0}{0}{0.4}{0}{2}                  % Node 2 (0% green = 100% orange)
\pienodewithanchor{n3t1}{2}{1}{0.4}{250}{3}                % Node 3 (40% green)

\draw (n1t1) -- (n3t1);
\draw (n2t1) -- (n3t1);
%\node at (1,3.2) {$t = 1$};

% --- t = 2 ---
\pienodewithanchor{n1t2}{4}{2}{0.4}{300}{1}                % Node 1 (90% green)
\pienodewithanchor{n2t2}{4}{0}{0.4}{0}{2}                  % Node 2 (100% orange)
\pienodewithanchor{n3t2}{6}{1}{0.4}{190}{3}                % Node 3 (60% green)

\draw (n1t2) -- (n2t2);
\draw (n2t2) -- (n3t2);
%\node at (5,3.2) {$t = 2$};

% --- t = 3 ---
\pienodewithanchor{n1t3}{8}{2}{0.4}{335}{1}                % Node 1 (90% green)
\pienodewithanchor{n2t3}{8}{0}{0.4}{30}{2}                  % Node 2 (100% orange)
\pienodewithanchor{n3t3}{10}{1}{0.4}{60}{3}               % Node 3 (90% green)
% Fully connected
\draw (n1t3) -- (n2t3);
\draw (n2t3) -- (n1t3);
\draw (n1t3) -- (n3t3);
\draw (n2t3) -- (n3t3);
\draw (n3t3) -- (n1t3);
\draw (n3t3) -- (n2t3);
%\node at (9,3.2) {$t = 3$};
\end{tikzpicture}

%% file: JASA_paper/latent_Markov_full.tex
\begin{tikzpicture}[
    >=Latex,
    every node/.style={font=\large},
    plate/.style={draw,very thick,rounded corners=2pt,minimum width=3.2cm,minimum height=4.9cm},
    var/.style={inner sep=1.5pt,anchor=west},
    ybox/.style={draw,very thick,minimum width=1.8cm,minimum height=1.1cm,align=center},
    lab/.style={font=\normalsize,inner sep=1pt},
    arr/.style={->,line width=.9pt},
    dotdot/.style={font=\Large}
]

% x positions for time slices
\def\xA{0}     % t-1
\def\xB{6.0}   % t
\def\xC{12.0}  % t+1

% Plate rectangles
\node[plate] (PA) at (\xA,0) {};
\node[plate] (PB) at (\xB,0) {};
\node[plate] (PC) at (\xC,0) {};

% Labels on top of each plate (invisible 
% Y coordinates inside a plate
\def\yTop{1.5}
\def\yMid{0.0}
\def\yBot{-1.7}

% Left plate contents (t-1)
\node[var] (Z1A) at ([xshift=-1.0cm]PA.center |- 0,\yTop) {$W^{(t-1)}_1$};
\node[var] (Z2A) at ([xshift=-1.0cm]PA.center |- 0,\yMid) {$W^{(t-1)}_2$};
\node[var] (ZNA) at ([xshift=-1.0cm]PA.center |- 0,\yBot) {$W^{(t-1)}_p$};
\node[dotdot] at ([xshift=-1.0cm]PA.center |- 0,{-0.9}) {$\vdots$};

% Middle plate contents (t)
\node[var] (Z1B) at ([xshift=-1.0cm]PB.center |- 0,\yTop) {$W^{(t)}_1$};
\node[var] (Z2B) at ([xshift=-1.0cm]PB.center |- 0,\yMid) {$W^{(t)}_2$};
\node[var] (ZNB) at ([xshift=-1.0cm]PB.center |- 0,\yBot) {$W^{(t)}_p$};
\node[dotdot] at ([xshift=-1.0cm]PB.center |- 0,{-0.9}) {$\vdots$};

% Right plate contents (t+1)
\node[var] (Z1C) at ([xshift=-1.0cm]PC.center |- 0,\yTop) {$W^{(t+1)}_1$};
\node[var] (Z2C) at ([xshift=-1.0cm]PC.center |- 0,\yMid) {$W^{(t+1)}_2$};
\node[var] (ZNC) at ([xshift=-1.0cm]PC.center |- 0,\yBot) {$W^{(t+1)}_p$};
\node[dotdot] at ([xshift=-1.0cm]PC.center |- 0,{-0.9}) {$\vdots$};

% Y boxes underneath each plate
\node[ybox] (YA) at (\xA,-4.3) {${N}^{(t-1)}$};
\node[ybox] (YB) at (\xB,-4.3) {${N}^{(t)}$};
\node[ybox] (YC) at (\xC,-4.3) {${N}^{(t+1)}$};

% Arrows Z -> Y (with phi^t labels)
\draw[arr] (PA.south)++(0,-0) -- (YA.north) node[midway,left=1mm,lab] {};
\draw[arr] (PB.south)++(0,-0) -- (YB.north) node[midway,left=1mm,lab] {};
\draw[arr] (PC.south)++(0,-0) -- (YC.north) node[midway,left=1mm,lab] {};
% Actually draw to the matching y-levels:
\draw[arr] (Z1A.east) -- ([xshift=-0.6cm]Z1B.west) node[midway,above,lab] {$\Gamma_1^{(t-1)}$};
\draw[arr] (Z2A.east) -- ([xshift=-0.6cm]Z2B.west) node[midway,above,lab] {$\Gamma_2^{(t-1)}$};
\draw[arr] (ZNA.east) -- ([xshift=-0.6cm]ZNB.west) node[midway,above,lab] {$\Gamma_p^{(t-1)}$};

\draw[arr] (Z1B.east) -- ([xshift=-0.6cm]Z1C.west) node[midway,above,lab] {$\Gamma_1^{(t)}$};
\draw[arr] (Z2B.east) -- ([xshift=-0.6cm]Z2C.west) node[midway,above,lab] {$\Gamma_2^{(t)}$};
\draw[arr] (ZNB.east) -- ([xshift=-0.6cm]ZNC.west) node[midway,above,lab] {$\Gamma_p^{(t)}$};
% Bottom row continuation dots
\node[dotdot] at ([yshift=-0.2cm]YA.south) {};
\node[dotdot] at ([yshift=-0.2cm]YB.south) {};
\node[dotdot] at ([yshift=-0.2cm]YC.south) {};
\end{tikzpicture}

%% file: JASA_paper/latent_markov_k.tex
\begin{tikzpicture}[
    % Global style settings
    >=Stealth, % Arrow tip style
    % Custom styles
    % INCREASED MINIMUM SIZE to 1.8cm to ensure all circles are the same size
    mathnode/.style={circle, draw, minimum size=1.4cm, inner sep=2pt, font=\bfseries}, 
    specialnode/.style={circle, draw=black, very thick, minimum size=1.3cm, font=\bfseries}, 
    textonly/.style={inner sep=0pt, font=\small}
]

% 1. Top Input Node - Coordinates fixed to align centrally
\node[mathnode] (AB) at (1.8, 3.8) {$a, b$};

% 2. B-nodes and Gamma nodes on the SAME HORIZONTAL LINE (y=0)
\node[mathnode] (Bt_1) at (-6, 0) {$B_{k}^{(t-1)}$}; % (-6, 0)
\node[mathnode, right=2.5cm of Bt_1] (Gamma_t_1) {$\Gamma_{k}^{(t-1)}$}; % (-3.5, 0)
\node[mathnode, right=2.5cm of Gamma_t_1] (Bt) {$B_{k}^{(t)}$}; % (-1.0, 0)
\node[mathnode, right=2.5cm of Bt] (Gamma_t) {$\Gamma_{k}^{(t)}$}; % (1.5, 0)
\node[mathnode, right=2.5cm of Gamma_t] (Bt1) {$B_{k}^{(t+1)}$}; % (4.0, 0)

% 4. W-nodes (Observations) - Below the B nodes (y=-3.0)
\node[mathnode, below=2.2cm of Bt_1] (W_t_1) {$W_{k}^{(t-1)}$};
\node[mathnode, below=2.2cm of Bt] (W_t) {$W_{k}^{(t)}$};
\node[mathnode, below=2.2cm of Bt1] (W_t1) {$W_{k}^{(t+1)}$};

% 5. Central W_0 node and Initialization
% NEW POSITION: Y-coordinate changed from 1.5 to 0.7
\node[mathnode] (psi) at (4, 1.8) {$\psi$};
% Input node is now a circle using the 'mathnode' style
\node[mathnode] (W0) at (4, -1.5) {$W_{0}$};
% Input node is now a circle using the 'mathnode' style
\node[mathnode, right=1cm of W0] (InputW0) {$\alpha, \sigma, \tau$};

% --- Draw Edges (Arrows) ---

% Center arrow (a,b) to B_k^t (Straight)
\draw[->] (AB) -- (Bt.north);

% Left arrow (a,b) to B_k^(t-1) (STRAIGHT DIAGONAL)
\draw[->] (AB.south west) -- (Bt_1.north);

% Right arrow (a,b) to B_k^(t+1) (STRAIGHT DIAGONAL)
\draw[->] (AB.south east) -- (Bt1.north);

% B-Gamma-B sequence (Horizontal Flow)
\draw[->] (Bt_1) -- (Gamma_t_1);
\draw[->] (Gamma_t_1) -- (Bt);
\draw[->] (Bt) -- (Gamma_t);
\draw[->] (Gamma_t) -- (Bt1);

% B-nodes to W-nodes (Downward Flow)
\draw[->] (Bt_1) -- (W_t_1);
\draw[->] (Bt) -- (W_t);
\draw[->] (Bt1) -- (W_t1);

% W_0 to W-nodes (DOWNWARD FLOW)
\draw[->] (W0) -- (W_t_1);
\draw[->] (W0) -- (W_t);
\draw[->] (W0) -- (W_t1);

% psi to Gmma-nodes (DOWNWARD FLOW)
\draw[->] (psi) -- (Gamma_t_1);
\draw[->] (psi) -- (Gamma_t);

% Input to W_0 (Upward Flow)
\draw[->] (InputW0) -- (W0);
\end{tikzpicture}